\documentclass[onecolumn,noshowpacs,nofootinbib,11pt]{revtex4-1}

\usepackage{float,xcolor,upgreek,yfonts,tikz}
\usepackage{color}
\usepackage{graphicx,bigints}
\usepackage{graphicx,float}\usepackage[all]{xy}
\usepackage{amsmath,upgreek}

   \usepackage{caption}
   
 \usepackage{booktabs,makecell}
\usepackage{subcaption}
\usepackage{amssymb}

\usepackage{alphabeta}
 \usepackage{cancel}
\usepackage{dcolumn}
\usepackage{textgreek}
\usepackage{multirow} 
\usepackage{units}
\usepackage{overpic}

\newcommand{\beq}{\begin{eqnarray}}
\newcommand{\eeq}{\end{eqnarray}}
\newcommand{\be}{\begin{equation}}
\newcommand{\ee}{\end{equation}}

\newcommand{\bea}{\begin{eqnarray}}
\newcommand{\eea}{\end{eqnarray}}

\newcommand{\ba}{\begin{eqnarray}}
\newcommand{\ea}{\end{eqnarray}}
\usepackage[colorlinks,hyperindex,unicode]{hyperref}
\definecolor{green1}{RGB}{0,128,0} 
\hypersetup{hidelinks,backref=true,pagebackref=true,hyperindex=true,colorlinks=true,breaklinks=true,urlcolor= blue}
\hypersetup{%
  colorlinks = true,
  linkcolor  = blue,
  citecolor = cyan,
}
\usepackage{bookmark,textgreek}
\usepackage{hyperref,color,xcolor}
\hypersetup{hidelinks,hyperindex=true,colorlinks=true,breaklinks=true,urlcolor= blue}
\hypersetup{%
  colorlinks = true,
  linkcolor  = blue
}
\newcommand\orcidcasadio{{\href{https://orcid.org/0000-0002-1330-7787}{\orcidicon}}}

\newcommand\orcidroldao{{\href{https://orcid.org/0000-0003-3978-532X}{\orcidicon}}}
\newcommand{\orcidicon}{%
	\begin{tikzpicture}
	\draw[lime, fill=lime] (0,0)
		circle [radius=0.16]
		node[white] {{\fontfamily{qag}\selectfont \tiny ID}};
	\draw[white, fill=white] (-0.0625,0.095)
		circle [radius=0.007];
	\end{tikzpicture}	\hspace{-2mm}
}

\begin{document}
\title{Axion stars in MGD background}

\author{R.~Casadio\orcidcasadio}
\affiliation{Dipartimento di Fisica e Astronomia, Universit\`a di Bologna, via Irnerio~46, 40126 Bologna, Italy}
\affiliation{I.N.F.N., Sezione di Bologna, I.S.~FLAG, viale B.~Pichat~6/2, 40127 Bologna, Italy}
\email{casadio@bo.infn.it}

\author{R. da Rocha\orcidroldao}
\affiliation{Center of Mathematics, Federal University of ABC, 09210-580, Santo Andr\'e, Brazil}
\email{roldao.rocha@ufabc.edu.br}
\affiliation{Dipartimento di Fisica e Astronomia, Universit\`a di Bologna, via Irnerio~46, 40126 Bologna, Italy}

\medbreak
\begin{abstract} 

 The minimal geometric deformation (MGD) paradigm is here employed to survey axion stars on fluid branes. The finite value of the brane tension provides beyond-general relativity corrections to the density, compactness, radius, and asymptotic limit of the gravitational mass function of axion stars, in a MGD background. The brane tension also enhances the effective range and magnitude of the axion field coupled to gravity.  MGD axion stars are compatible to mini-massive compact halo objects for almost all the observational range of brane tension, however, a narrow range allows MGD axion star densities large enough to produce stimulated decays of the axion to photons,  with no analogy in the general-relativistic (GR) limit.  Besides, the gravitational mass and the density of MGD axion stars are shown to be up to four orders of magnitude larger than the GR axion stars, being also less  sensitive to tidal disruption events under collision with neutron stars, for lower values of the fluid brane tension. 

\end{abstract}



\maketitle
\section{Introduction}

The experimental measurement of gravitational-wave (GW) signatures radiated from the final stages of neutron star binary merging constitutes one of the most relevant results in fundamental physics \cite{LIGOScientific:2017vwq}. 
In the strong regime of gravity, general-relativistic solutions of Einstein's equations and their generalizations may be experimentally detected by the latest observations mainly at LIGO, Chandra, eLISA, Virgo, GEO600, TAMA 300, and KAGRA detectors, as well as the next generation, including the Advanced LIGO Plus, Advanced Virgo Plus, and the Einstein telescope. These detectors can thoroughly address extended models of gravity, whose solutions of Einstein's effective equations describe coalescent binary systems composed of stars, or even merging black holes, thus emitting GW-radiation in the endpoint stages of collision after spiraling in against each other. The gravitational decoupling (GD) of Einstein's equations has been successfully extending general relativity (GR) and has been modeling a multitude of self-gravitating compact stellar configurations. Anisotropic stars arise in a very natural way in the GD apparatus, yielding the possibility of obtaining the state-of-the-art of analytical solutions of Einstein's equations, when more general forms of the energy-momentum tensor are employed \cite{Ovalle:2017fgl,Ovalle:2019qyi,Ovalle:2013vna,Ovalle:2018vmg,Ovalle:2018gic}. 
The GD mechanism  comprises the original minimal geometrical deformation (MGD) \cite{Casadio:2012rf,Ovalle:2007bn,Ovalle:2018ans}, which formulates the description of compact stars and black hole solutions of Einstein's equations on fluid branes, with finite brane tension \cite{Antoniadis:1998ig,daRocha:2012pt,Abdalla:2009pg}. GR is the very limit of 
the fluid brane setup, when the brane describing our universe is ideally rigid, corresponding to an infinite value of the brane tension. 
When the GD is implemented into the so-called analytical seed solutions of Einstein's equations,  all sources generating the gravitational field are decomposed into two parts. The first one includes a GR solution, whereas the second piece refers to a complementary source, which can carry any type of charge, including tidal and gauge ones, hairy fields of some physical origin, as well as any other source which plays specific roles in extended models of gravity. Quasinormal modes radiated from hairy GD solutions were recently addressed in Ref. \cite{Cavalcanti:2022cga}. 
The GD methods have been comprehensively employed to engender extended solutions reporting an exhaustive catalog of stellar configurations \cite{Estrada:2019aeh,Gabbanelli:2019txr,Leon:2023nbj,daRocha:2020jdj,Avalos:2023ywb,Contreras:2022vec,Avalos:2023jeh,Maurya:2021zvb,Maurya:2021mqx,Singh:2021iwv,Maurya:2021tca,Maurya:2023szc,Jasim:2021kga,Singh:2020bdv,Cavalcanti:2016mbe,Ramos:2021drk,Casadio:2019usg,Rincon:2019jal,Tello-Ortiz:2019gcl,Morales:2018urp,Panotopoulos:2018law,Singh:2019ktp,Jasim:2023ehu,Maurya:2020djz}, which in particular well describe an anisotropic star that was recently observed  \cite{Gabbanelli:2018bhs,PerezGraterol:2018eut,Heras:2018cpz,Torres:2019mee,Hensh:2019rtb,Contreras:2019iwm,Tello-Ortiz:2021kxg,Sharif:2020vvk,Contreras:2019mhf}. Not only restricted to the gravity sector of AdS/CFT, the quantum holographic entanglement entropy was also studied in the GD context \cite{daRocha:2019pla}. 
GD-anisotropic quark and neutron stars were scrutinized in Refs. \cite{Contreras:2021xkf,Sharif:2020lbt,Maurya:2020ebd}, whereas GD-black holes with hair were also reported in Refs. \cite{Ovalle:2020kpd,Ovalle:2021jzf,Contreras:2021yxe,Meert:2020sqv}.

The paradigm of formulating dark matter (DM) dominating ordinary matter in galaxies is based upon precise observational data from measuring the CMB by  Planck Collaboration 
\cite{Planck:2018vyg}. Despite fruitful observational data confirming the existence of DM, its very nature remains concealed. Even though diverse particles   have been proposed as the ruling component of DM, hardly any particle candidate can properly present the properties of DM. The axion is an  
exception and plays the role of a DM prime candidate. In this context, taking into account the spontaneous breaking of the Peccei--Quinn (PQ) symmetry after inflation in the early universe, axion miniclusters can have originated \cite{Hogan:1988mp}.  
The axion is the Nambu--Goldstone pseudoscalar  boson, generated in the spontaneous breaking of the U${}_{\textsc{PQ}}$(1) global symmetry.
 The PQ symmetry was originally introduced to report the tininess of the strong CP violation that occurs in the QCD context. The axion can couple to two real photons, being therefore detected by its conversion into a photon, in a strong enough magnetic field. Since stellar distributions usually present strong magnetic fields, axions may be originated in their inner core, in the course of the cooling process,  and can have annihilated into photons.  
 Besides, axion can also form miniclusters. Also owing to the gravitational cooling, some regions of the axion minicluster can develop themselves colder than other portions, when axion particles are ejected. This process leads to the formation of self-gravitating axion stars \cite{Seidel:1993zk,Levkov:2018kau,Bai:2021nrs}. There is another way for axion miniclusters to turn into compact axion stars. Due to the (attractive) self-interaction, some nonlinear effects can yield very dense axion miniclusters. If such density peaks are high enough, coherent axion fields can amalgamate into a self-gravitating system comprising axion stars. Refs. \cite{Barranco:2010ib,Barranco:2012ur} addressed the problem of describing DM with axions, exploring the resulting astrophysical self-gravitating objects made of axions, in the general-relativistic case. Axion stellar configurations are bound together by way of equilibrium among gravitational attraction, kinetic pressure, and an intricate self-interaction \cite{Eby:2020ply}. 
Ref. \cite{Iwazaki:2022bur} proposed observational signatures of 
radiation bursts when axion stars collide with galaxies, whereas Ref. \cite{Witte:2022cjj} also put forward the possibility of observing radio signals from axions miniclusters and axion stars merging with neutron stars \cite{Witte:2022cjj}. 
Other spacetimes with axion fields were scrutinized in Refs. \cite{Casadio:1997sd,Casadio:1996rp}. 

In this work, we address the possibility of describing DM with axions, exploring  astrophysical axion stars in an MGD background, in the membrane paradigm of AdS/CFT. Using an AdS bulk of codimension one with respect to its 4-dimensional boundary describing our universe, with a finite brane tension, is quite natural in several scenarios. Axion particles are introduced in a beyond-standard model context, being ubiquitous in string theory compactifications. In this scenario,  the axion can be characterized by a Kaluza--Klein pseudoscalar, associated with (non-trivial) cycles in the compactified geometry \cite{Svrcek:2006yi}. 

 A weak MGD background will be assumed to solve the Einstein--Klein--Gordon (EKG) equations, coupling the axion to gravity.  The solutions of Einstein's field equations describing the gravitational sector in an MGD background, coupled with the Klein--Gordon equation with the axion potential, will 
 produce an effective static spherically symmetric metric. The asymptotic value of gravitational masses, the radii, the densities, and the compactnesses of 
 MGD axion stars will be computed and shown to be a function of the brane tension. The gravitational mass, the density,  and the compactness of self-gravitating systems composed of axions will be shown to be magnified,  when compared to the general-relativistic scenario, for a considerable range of the MGD parameter which encodes the brane tension. A couple of relevant results are obtained, with no analogy in GR. The first one consists of obtaining the typical densities of MGD axion stars. Contrary to the GR case, where the axion stars have densities of around 4 orders of magnitude smaller than neutron stars, MGD axion stars  can reach magnitudes that approach typical densities of neutron stars,  for a range of the brane tension lying into the latest allowed observational bounds. It allows the detection of observational signatures of collisions between MGD axion stars and neutron stars, which are completely different from the GR axion stars.  By the fact that neutron stars are surrounded by strong magnetic fields, 
photons are supposed to be ejected by the collision process with axions \cite{Seidel:1993zk,Tkachev:1987cd}. If the plasma constituted by photons near the neutron star has the same order of the axion mass, the axion conversion into a photon is then coherent \cite{Bai:2021nrs}. The photons that are emitted have typical radio-wave frequencies and might be detected by  ground-based telescopes, such as the ones in the Square Kilometre Array  and the Green Bank Observatory. When an axion star crosses the way of a neutron star, if they are nearer than a given radius, the tidal force induced by the neutron star sets off stronger than the self-gravity of the axion star. 
Such kind of compact object is called a diluted axion star, which plays the role of a Bose--Einstein-like condensate, whose gravity balances quantum pressure.
The dilute axion star can be thoroughly fragmented by tidal forces, before attaining the radius for which the plasma of photons has the same mass as the axion. A 2-body tidal capture mechanism can be then investigated for MGD axion stars. 

Even in the GR case, the study of collisions of axion stars to neutron stars is still incipient \cite{Du:2018qor}. Here we want to shed new light on this topic, proposing corrections to the asymptotic value of gravitational masses, the radii, the densities, and the compactnesses of  MGD axion stars. Since MGD axion will be shown to present 
typical masses and densities that can reach 4 orders of magnitude larger than GR axion stars, for a given range of brane tension, the maximum distance 
for which the axion star undergoes disruption event and the percentage of axions that can be converted into photons, across the collision event to neutron stars, will be quite different. When one takes into account phenomenologically feasible values of the axion mass and the axion decay constant, we will also show that there are ranges of the brane tension allowing stimulated decay of axions into photons, implying that the final stage of the collapse process induced by gravitational cooling is a flash of photons, which has no parallel in the general-relativistic limit. 

This paper is organized as follows: Sec. \ref{MGD} introduces the MGD method, yielding analytical solutions of Einstein's field equations on the brane, modeling realistic compact stellar distributions in a membrane paradigm of AdS/CFT, with finite brane tension. 
 In Sec. \ref{EKG}, the quantized axion field, described by the Klein--Gordon equation with an appropriate potential, is coupled to Einstein's equations. The resulting solutions of the EKG coupled system of equations produce an effective static spherically symmetric metric. The asymptotic value of gravitational masses, the densities, the radii, and the compactnesses of 
 the compact self-gravitating system of axions are analyzed for several values of the parameter regulating the MGD solutions. Sec. \ref{main} is dedicated to taking phenomenologically feasible values of the axion mass and the axion decay constant, scrutinizing axion stars in an MGD background, in a setup compatible with mini-massive compact halo objects. We show that there are ranges of the brane tension allowing stimulated decay of axions into photons implying that the final stage of the collapse process induced by gravitational cooling is a flash of photons, which has no parallel in the general-relativistic limit. 
 Several other physical features of MGD axion stars are addressed, with important corrections to the general-relativistic limit. One of the main relevant results consists of proposing MGD axion stars with masses and densities that make them less sensitive to tidal disruption, in collisions with neutron stars, for a certain range of the brane tension. The maximum distance beyond which MGD axion stars undergo tidal disruptive events is computed for several values of the central value of the axion field and is shown to be an increasing function of the MGD parameter. With it, we show that MGD axion stars are less  sensitive to tidal disruption effects, as the brane tension decreases.
Sec. \ref{cppp} is devoted to the conclusions, further discussion, and perspectives.

\section{The MGD protocol}
\label{MGD}

The MGD is naturally developed in the membrane paradigm of AdS/CFT, wherein a finite value of the brane tension mimics the energy density, $\upzeta$, of the brane. The brane tension and both the running cosmological parameters on the brane and in the bulk are tied together by fine-tuning \cite{maartens}. 
Any physical system having energy satisfying $\upzeta\gg E$ perceive neither bulk effects nor self-gravity, allowing GR to be recovered as the ideally rigid brane ($\upzeta\to\infty$) limit. However, for $\upzeta\lesssim E$, finite brane tension values can yield new physical possibilities.  The most recent and accurate brane tension bound, $\upzeta \gtrsim 2.832\times10^{-6} \,{\rm GeV^4}$, was obtained, in the context of the MGD  \cite{Fernandes-Silva:2019fez}. 

 The MGD algorithm has been extensively utilized for constructing new analytical solutions of Einstein's equations, encompassing new aspects of extended models of gravity to classical GR solutions, when a fluid membrane setup is taken into account \cite{darkstars}. In the brane-world scenario, the 4-dimensional membrane, which portrays the universe we live in, is usually embedded into a codimension one AdS bulk space. Therefore the Gauss--Codazzi equations  link together the induced metric and the extrinsic curvature of the brane, considered as a submanifold of the AdS bulk. In this scenario, the Riemann tensor of the AdS bulk   is split into the sum of the Riemann tensor of the brane and quadratic terms of the extrinsic curvature. Einstein's equations on the brane are given by  
\begin{equation}
\label{5d4d}
\mathtt{R}_{\mu\nu} - \frac12\mathtt{R}g_{\mu\nu}
=\Uplambda_4 g_{\mu\nu}+\mathtt{T}_{\mu\nu},
\end{equation} where $\mathtt{R}$ stands for the Ricci scalar, $8\pi G_4=1$ will be adopted throughout this work, $G_4$ is the brane Newton constant, and $\mathtt{R}_{\mu\nu}$ denotes the Ricci tensor, whereas $\Uplambda_4$ is brane cosmological running parameter. The energy-momentum tensor in Eq. (\ref{5d4d}) is usually decomposed as  \cite{GCGR}
\beq
\mathtt{T}_{\mu\nu}
=
T_{\mu\nu}+{\scalebox{0.98}{$\mathtt{E}$}}_{\mu\nu}+\upzeta^{-1} \mathtt{S}_{\mu\nu}.\label{tmunu}\eeq 
The $T_{\mu\nu}$ term is the energy-momentum tensor representing ordinary matter, eventually also describing dark matter. The term ${\scalebox{0.98}{$\mathtt{E}$}}_{\mu\nu}$ is the projection of the Weyl tensor along the brane directions and is a function inversely proportional to the brane tension. The tensor $\mathtt{S}_{\mu\nu}$ reads  	
\begin{eqnarray}\label{smunu}
\!\!\!\mathtt{S}_{\mu\nu} = \frac{1}{3}T\,T_{\mu\nu}-T_{\mu\upsilon }T^\upsilon_{\ \nu} + \frac{1}{6}\Big[3T^{\alpha\sigma}T_{\alpha\sigma} - T^2\Big]\,g_{\mu\nu},
\end{eqnarray}
\noindent where $T=T_{\mu\nu}g^{\mu\nu}$ \cite{GCGR,maartens,Antoniadis:2011bi}.
The electric part of the Weyl tensor,  
\begin{eqnarray}
\!\!\!\!\!\!\!\!{\scalebox{0.98}{$\mathtt{E}$}}_{\mu\nu} \!&=&\!-\frac{1}{\upzeta}\!\left[ {\scalebox{0.98}{$\mathtt{U}$}}\!\left(\!{\scalebox{0.98}{$\mathtt{u}$}}_\mu {\scalebox{0.98}{$\mathtt{u}$}}_\nu \!+\! \frac{1}{3}\mathtt{h}_{\mu\nu}\!\right)+ {\scalebox{0.98}{$\mathtt{Q}$}}_{(\mu} {\scalebox{0.98}{$\mathtt{u}$}}_{\nu)}\!+\!{\scalebox{0.98}{$\mathtt{P}$}}_{\mu\nu}\right], \label{A4}
\end{eqnarray}
\noindent characterizes a Weyl-type fluid, for $\mathtt{h}_{\mu\nu}$ emulating the  induced metric projecting any quantity along the direction that is orthogonal to the velocity field ${\scalebox{0.98}{$\mathtt{u}$}}^\upsilon$ regarding the Weyl fluid flow. Also, the dark radiation, the anisotropic energy-momentum tensor, and the   flux of energy can be described by functions of the brane tension, respectively  as 
\beq\label{drt}
{\scalebox{0.98}{$\mathtt{U}$}}&=&-\upzeta\,{\scalebox{0.98}{$\mathtt{E}$}}_{\mu\nu} {\scalebox{0.98}{$\mathtt{u}$}}^\mu {\scalebox{0.98}{$\mathtt{u}$}}^\nu,\\\label{eed}
{\scalebox{0.98}{$\mathtt{P}$}}_{\mu\nu}&=&-\upzeta\,\left[\mathtt{h}_{(\mu}^{\;\rho}\mathtt{h}_{\nu)}^{\;\sigma}-\frac13 \mathtt{h}^{\rho\sigma}\mathtt{h}_{\mu\nu}\right]{\scalebox{0.98}{$\mathtt{E}$}}_{\rho\sigma},\\\label{eee}
{\scalebox{0.98}{$\mathtt{Q}$}}_\mu &=& -\upzeta\, \mathtt{h}^{\rho\sigma}{\scalebox{0.98}{$\mathtt{E}$}}_{\sigma \mu}{\scalebox{0.98}{$\mathtt{u}$}}_\rho.\eeq 
One can therefore realize the equations governing the gravitational sector on the brane from holographic AdS/CFT, since the electric component of the Weyl tensor can be expressed, in the linear order, as the energy-momentum tensor of CFT living on the brane \cite{Shiromizu:2001jm}. Going further to higher-order terms yields a relationship between the tensor in (\ref{smunu}) and the trace anomaly of CFT \cite{Meert:2020sqv}. 

Denote hereon by $\ell_p=\sqrt{\frac{G_4\hbar}{c^3}}$ the Planck length and by $G_5$ the bulk Newton running parameter. It was shown to be related
to $G_4$ by the Planck length, as  $G_5 = G_4\ell_p$ \cite{maartens}. Denoting by $\upkappa_5 = 8\pi G_5/c^2$, the 4-dimensional and 5-dimensional cosmological running parameters are fined tuned to the brane tension by the expression \cite{Randall:1999vf}
\be
\Uplambda_4=\frac{\upkappa_5^{2}}{2}\Big(\Uplambda_{5}+\frac{1}{6}\upkappa_5^{2}\upzeta^{2}\Big).\label{fine}\ee 
Eq. (\ref{fine}), together with the fact that the 4-dimensional coupling constant $\upkappa_4 = 8\pi G_4/c^2$ is related to $\upkappa_5$ by $\upkappa^2_{4}=\frac{1}{6}\upzeta\upkappa^4_5$,  yields  \cite{maartens} 
\beq\label{uuup}
\Uplambda_{5}=-\frac{\sqrt{6}}{6}\upkappa_4\upzeta^{3/2}.
\eeq
what complies with an AdS bulk. The finite brane tension is related to the 5-dimensional Planck mass, $m_{p5}$, by 
$
\upzeta\approx \pi{\mathtt{r}}\sqrt{{-\Uplambda_5}/{24m_{p5}^3}}, 
$ where $
\mathtt{r}\approx 1\,\mu{\rm m},
$  With the expression of the extrinsic curvature  \cite{maartens}
\beq
K_{\mu\nu}=-\frac{1}{2}\upkappa^2_5\left[T_{\mu\nu}+\frac13\left(\upzeta-T\right)g_{\mu\nu}\right],
\eeq
the electric part of the Weyl tensor can be alternatively expressed by \cite{maartens}
\beq
{\scalebox{0.98}{$\mathtt{E}$}}_{\mu\nu}=-\frac{\Uplambda_{5}}{6}g_{\mu\nu}-\partial_z K_{\mu\nu}+K_\mu^{\;\rho}K_{\rho\nu},
\eeq
where $z$ denotes the Gaussian coordinate along the bulk.

Compact stars are solutions of Einstein's effective field equations (\ref{5d4d}), with static and spherically symmetric metric 
\begin{equation}\label{abr}
ds^{2} = a(r) dt^{2} - \frac{1}{b(r)}dr^{2} - r^{2}d\Omega^2, 
\end{equation} where $d\Omega^2$ is the solid angle. 
In this context, the tensor fields in Eqs. (\ref{eed}, \ref{eee}) take a simplified expression, respectively given by ${\scalebox{0.98}{$\mathtt{P}$}}_{\mu\nu} = {\scalebox{0.98}{$\mathtt{P}$}}({\scalebox{0.98}{$\mathtt{u}$}}_\mu {\scalebox{0.98}{$\mathtt{u}$}}_\nu+ \mathtt{h}_{\mu\nu}/3)$ and ${\scalebox{0.98}{$\mathtt{Q}$}}_\mu = 0$, where ${\scalebox{0.98}{$\mathtt{P}$}}=g_{\mu\nu}{\scalebox{0.98}{$\mathtt{P}$}}^{\mu\nu}$. The brane  energy-momentum tensor can be prescribed by a perfect fluid one, encoding different particles and fields confined to the brane, as  \beq
T_{\mu\nu} = (\upepsilon + p) {\scalebox{0.98}{$\mathtt{u}$}}_\mu {\scalebox{0.98}{$\mathtt{u}$}}_\nu- pg_{\mu\nu},\label{pfc}\eeq with ${\scalebox{0.98}{$\mathtt{u}$}}_\mu=\sqrt{b(r)}\delta_\mu^0$.
Now the MGD-decoupling method will be shown to yield analytical solutions of Einstein's equations on the brane (\ref{5d4d}, \ref{tmunu}). These solutions can  realistically represent compact stars on fluid branes   \cite{Ovalle:2017fgl,Casadio:2012rf,darkstars}.

The Einstein's equations on the brane \eqref{5d4d} denoting by ${\scalebox{0.98}{$\mathtt{G}$}}_{\mu\nu} = R_{\mu\nu} - \frac12\mathtt{R}g_{\mu\nu}$ can read \begin{subequations} 
\beq\label{eqw}
\label{usual} 
\!\!\!\!\!\!\!\!\!\!\!\!\!\!\!\!\!\!b(r)\;&=&1-\frac{1}{r}\int_0^r \!\mathtt{r}^2\upepsilon(\mathtt{r})\left[1+\frac{1}{2\,\upzeta}\!\!\left(\upepsilon(\mathtt{r})\!+\!\frac{3{\scalebox{0.98}{$\mathtt{U}$}}(\mathtt{r})}{8}\right)\!\right]d\mathtt{r},\\
\label{pp}
\!\!\!\!\!\!\!\!\!\!\!\!\!\!\!\!\!\!\!{\mathtt{
P}}(r)&=&\frac{\upzeta}{6}\left[{\scalebox{0.98}{$\mathtt{G}$}}_1^{\,1}(r)-{\scalebox{0.98}{$\mathtt{G}$}}_2^{\,2}(r)\right],\\
\label{uu}
\!\!\!\!\!\!\!\!\!\!\!\!\!\!\!\!\!\!\!{{\scalebox{0.98}{$\mathtt{U}$}}}(r)&\!=\!&-16\upepsilon(r)\left(\frac{\upepsilon(r)}{2}\!+\!\frac{16}{3}
p(r)\right)\!+\!\upzeta\left(2{\scalebox{0.98}{$\mathtt{G}$}}_2^{\,2}(r)+{\scalebox{0.98}{$\mathtt{G}$}}_1^{\,1}(r)\right)\!-\!16p(r) \upzeta,\\
\label{con1}
\!\!\!\!\!\!\!\!\!\!\!\!\!\!\!\!\!\!\!p^\prime(r)&=&-\frac{a^\prime(r)}{2a(r)}\left[\upepsilon(r)+p(r)\right],
\eeq
\end{subequations}
where
\begin{subequations} 
\beq
\label{g11}{\scalebox{0.98}{$\mathtt{G}$}}_1^{\,1}(r)&=&-\frac 1{r^2}+\frac{1}{b(r)}\left( \frac
1{r^2}+\frac{a'(r)}{ra(r)}\right),\\
\label{g22}{\scalebox{0.98}{$\mathtt{G}$}}_2^{\,2}(r)&\!=\!&\frac{1}{4b(r)}\left[ \frac{2a''(r)}{a(r)}-\frac{a'(r)b'(r)}{a(r)b(r)}+\frac{a^{\prime 2}(r)}{a^2(r)}+\!\frac{1}{r}\left(\frac{b'(r)}{b(r)}-\frac{a'(r)}{a(r)}\right)\right].
\eeq
\end{subequations}
 GR can be immediately recovered whenever  the rigid brane limit 
$\upzeta\to\infty$ is taken into account.

The effective density ($\check{\upepsilon}$), the radial pressure ($\check{p}_{r}$), and also the tangential pressure ($\check{p}_{\intercal}$), are respectively expressed as \cite{darkstars}
\beq
\check{\upepsilon}&=&\upepsilon +\strut \displaystyle\frac{1}{ 2\upzeta }\left({\upepsilon ^{2}}+{3 {\scalebox{0.98}{$\mathtt{U}$}}}\right) \,, \label{efecden}\\
\check{p}_{r}&=&p+\frac{1}{2\upzeta }\left({\upepsilon ^{2}}+2\upepsilon \,p+{{\scalebox{0.98}{$\mathtt{U}$}}}+{\scalebox{0.98}{$\mathtt{P}$}}\right) \,, \label{efecprera}\\
\check{p}_{\intercal}&=&p+\frac{1}{2\upzeta }\left({\upepsilon ^{2}}+2\upepsilon\, p+{{\scalebox{0.98}{$\mathtt{U}$}}}-{\scalebox{0.98}{$\mathtt{P}$}}\right).\label{efecpretan}
\eeq

Gravity living in the AdS bulk yields the MGD term, $\upxi(r)$, in the radial metric term
 \begin{eqnarray}
\label{edlrwssg}
b(r)
&=&
\mu(r)+\upxi(r)
\ ,
\end{eqnarray}
where  
\begin{eqnarray}\label{muuu}
\mu(r) = \begin{cases} 1-\frac{1}{r}\bigintss_{\,0}^r \,\upepsilon({\rm r}) \,{\rm r}^2\,d{\rm r}
\,,
&
\quad r\,\lesssim\,R\,,
\\
1-\frac{2\mathsf{M}(\upzeta)}{r}
\ ,
&
\quad r\gtrsim R,
\end{cases}
\end{eqnarray} for $R$ denoting the star radius. 
The mass function can be written as the sum of the GR mass function and  terms of order running with the inverse of the brane tension \cite{Casadio:2012rf}:
 \beq\mathsf{M}(\upzeta)=M_0+{\cal O}(\upzeta^{-1}).\label{oioi}\eeq 
 Eq. (\ref{muuu}) can be expressed in a more compact version, as 
\begin{eqnarray}\label{muuu1}
\mu(r) =1-\frac{2\mathsf{M}(\upzeta, r)}{r}.
\end{eqnarray}

The general solution of the coupled system of ODEs (\ref{eqw}) -- (\ref{con1}) can be calculated by replacing Eq. (\ref{uu}) into Eq. (\ref{usual}). This method implies that  
\begin{eqnarray}
\label{e1g}
&&\frac{1}{b}\left(\frac{{4r^2a^{\prime 2}+{4a^2}}}
{r^2a'+4ar}-\frac{b'}{b}\right)=\frac{4a}{r\left(a'r+4a\right)}\left[1+{4 a r}{\left(\frac{\upepsilon}{\upzeta}
(\upepsilon+3p)+\upepsilon-3p\right)}\right],
\end{eqnarray}
with 
\begin{eqnarray}
\label{primsol}
b(r)&=&-e^{-I(r)}\!\left(\bigintsss_{\,0}^r\frac{e^I}{4a\mathtt{r}}\left(a'\mathtt{r}+4a\right)\left[\left(\upepsilon\!-\!\frac{\upepsilon}{\upzeta}\left(\upepsilon+3
p\right)-3p\right)-\frac{2}{\mathtt{r}^2}\right]d\mathtt{r}+\upbeta(\upzeta)\right),
\end{eqnarray}
for $\upbeta=\upbeta(\upzeta)$ being an function that is inversely proportional to  $\upzeta$, such that its GR limit vanishes, $\lim_{\upzeta\to\infty}\upbeta(\upzeta)=0$, whereas  
 \beq
\label{I}
\!\!\!\!\!\!\!\!\!\!I(r)=
\bigintsss_{\,0}^r\left[\frac{1}{\mathtt{r}^2a'+{4\mathtt{r}}}a^{\prime 2}(\mathtt{r}^2-1)
+4\mathtt{r}\,a'+4a^2\right]
\,d\mathtt{r}
\,.
\eeq
In order to the MGD seed \eqref{edlrwssg} match Eq. (\ref{primsol}), one must require that   
\begin{eqnarray}\label{kapp}
\!\!\!\!\!\!\!\!\!\!\!\upxi(r)\!=\!{e^{-I(r)}\!\left[\upbeta\!+\!\!\int_0^r\frac{2\mathtt{r}ae^I}{r a'+4a}
\left(\mathtt{L}\!+\!\frac{\upepsilon}{\upzeta}\left(\upepsilon+3p\right)\right)\right]
d\mathtt{r}},
\end{eqnarray}
for 
\beq
\label{L} \!\!\!\!\!\!\mathtt{L}(r)\;&\!=\!&
\left[\mu\left(\frac{a^{\prime 2}}{a^2}-\frac{a^{\prime 2}}{a^2}+\frac{2a'}{ar}+\frac{1}{r^2}\right)+\mu'\left(\frac{a'}{2a}+\frac{1}{r}\right)-\!\frac{1}{r^2}\right]-3p,
\eeq
The geometric deformation $\upxi(r)$ in the vacuum, denoted by $\upxi^{\scalebox{0.6}{$\star$}}(r)$, is minimal and it can be immediately computed when Eq.~(\ref{kapp}) is constrained to $\mathtt{L}(r)=0$, yielding 
\begin{equation}
\label{def}
\upxi^{\scalebox{0.6}{$\star$}}(r)
=\upbeta(\upzeta)\,e^{-I(r)}
\ .
\end{equation}
The radial metric component in Eq.~\eqref{edlrwssg} then becomes
\begin{eqnarray}
\label{g11vaccum}
\frac{1}{b(r)}=
{1-\frac{2\mathsf{M}}{r}}+\upbeta(\upzeta)\,e^{-I}\, ,
\end{eqnarray}
When the Israel conditions are employed to match the outer and inner geometry, together with Eqs. (\ref{muuu}, \ref{muuu1}), for $r\lesssim R$ the MGD metric is given by   
\begin{equation}
ds^{2}
\!=\!
a_{\scalebox{0.6}{$-$}}(r)dt^{2}
-\frac{1}{1-\frac{2{\mathsf{M}}(r)}{r}}dr^2
-r^{2}d\Upomega^2,
\label{mgdmetric}
\end{equation}
for the effective gravitational mass   \cite{darkstars}
\begin{equation}
\label{effecmass}
{\mathsf{M}}(r)
=
{M_0}(r)-\frac{r}{2}{\upxi^{{\scalebox{0.6}{$\star$}}}(r)}, 
\end{equation}
whereas in the outer region, Eq. (\ref{drt}) and the trace of (\ref{eed}) respectively read 
\ba
\label{pp2}
\,{{\scalebox{0.98}{$\mathtt{P}$}}_{\scalebox{0.6}{+}}}{(r)}
&=&
\frac{{4{\mathsf{M}(r)}}{}-3r}{27 \upzeta r^4\left(1-\frac{3{\mathsf{M}(r)}}{2r}\right)^{\!2}}\,\upbeta(\upzeta),\\
{{\scalebox{0.98}{$\mathtt{U}$}}_{\scalebox{0.6}{+}}}{(r)}
&=&
\frac{{\mathsf{M}(r)}}{12 \upzeta r^4\left(1-\frac{3{\mathsf{M}(r)}}{2r}\right)^{\!2}}\,\upbeta(\upzeta).
\ea
As both $p(r)$ and $\upepsilon(r)$ vanish in the outer region of the MGD stellar distribution, its metric reads  \cite{Casadio:2012rf}
\begin{equation}
\label{genericext}ds^2\!=\!a_{\scalebox{0.6}{+}}(r) dt^2\!-\frac{dr^2}{1\!-\!\frac{2\mathsf{M}(r, \upzeta)}{r}\!-\!\upxi^{\scalebox{0.6}{$\star$}}(r)}+r^{2}d\Upomega^2.
\end{equation}
At the star surface, $r=R$, the Israel matching conditions yield  \cite{Casadio:2012rf}
\begin{eqnarray}
 \label{ffgeneric1}
a_{\scalebox{0.6}{$-$}}(R)&=&\exp\left[1-\frac{2{\mathsf{M}(R)}}{R}\right]
=a_{\scalebox{0.6}{$+$}}(R),\\
 \label{ffgeneric2}
{\mathsf{M}(R)-M_0} &=&\frac{R}{2}\left[\upxi^{\scalebox{0.6}{$\star$}}_{\scalebox{0.6}{+}}(R)-\upxi^{\scalebox{0.6}{$\star$}}_{\scalebox{0.6}{$-$}}(R)\right].
\end{eqnarray}
The  Schwarzschild-like solution, \beq
a_{\scalebox{.6}{Schwa}}(r)=\frac{1}{b_{\scalebox{.6}{Schwa}}(r)}
=
1-\frac{2{\mathsf{M}(r)}}{r},\eeq can be now superseded into Eq.~\eqref{def}, yielding the MGD term to be equal to  
\be
\label{defS}
\upxi^{\scalebox{0.6}{$\star$}}(r)=
-\frac{4(r-2\mathsf{M}(r))}{\left(2r-{3\mathsf{M}(r)}\right)}\,\upbeta(\upzeta).
\end{equation}
In addition, at the surface of the MGD star, it follows that $\upxi^{\scalebox{0.6}{$\star$}}(r)<0$. 
The function $\upbeta(\upzeta)$ can be also expressed as  
\begin{equation}
\label{betafinal3}
\upbeta(\upzeta)
=\frac{1}{2 \upzeta R}\left(\frac{2R-{3{M_0}}}{R-{2\mathsf{M}(R)}}\right).
\end{equation}
Therefore, for $r>R$, the metric endowing the spacetime surrounding the MGD star has the following expression:
\begin{subequations}
\ba
\label{nu}
\!\!\!\!\!\!a(r)
&=&
1-\frac{2\mathsf{M}(r)}{r}
\ ,
\\
\!\!\!\!\!\!b(r)
&=&
\left(1+\frac{2{\scalebox{0.98}{$\mathfrak{l}$}}}{2r-{3\mathsf{M}(r)}}\right)\left(1-\frac{2\mathsf{M}(r)}{r}\right),
\label{mu}
\ea
\end{subequations} 
where 
\begin{equation}
\label{Lk}
{\scalebox{0.98}{$\mathfrak{l}$}}={\scalebox{0.98}{$\mathfrak{l}$}}(\upzeta)=
\frac{1}{\upzeta}
\left(\frac{2R-3M_0}{2R-4M_0}\right)\frac{2R-3\mathsf{M}(R)}{2R-4\mathsf{M}(R)}
\end{equation} 
is the MGD parameter, which depends on the value of the brane tension. 
The GR limit of a rigid brane, $\upzeta\to\infty$, thus recovers the Schwarzschild metric. Gravitational lensing effects in the strong regime, read off the supermassive black hole at the center of the Milky Way, the Sgr $A^{\scalebox{0.6}{$\star$}}$, established the bound $|{\scalebox{0.98}{$\mathfrak{l}$}}|\lesssim6.370\times 10^{-2}\ {\rm m}$ for the MGD parameter \cite{Cavalcanti:2016mbe}.

\section{Axion field coupled to gravity in MGD background }\label{EKG}

Axions are usually introduced in beyond-Standard Model physics. In the low-energy regime, axion phenomenology is regulated by two energy scales, comprising the axion mass, $m_\mathfrak{a}$, and
the axion decay constant, $f_\mathfrak{a}$, set to the order bigger than the electroweak scale ${\displaystyle f_{a}\approx 0.246}$ TeV to ensure that the axion field behaves similarly to the Higgs field \cite{Braaten:2019knj}. Astrophysical and cosmological observations limit the range $10^{-6}$ eV $\lesssim m_\mathfrak{a}\lesssim 10^{-3}$ eV. 
In this way, the axion field can be an adequate candidate for describing the cold DM as well as it can form Bose--Einstein condensates. Axions can be described by (pseudo)-Goldstone
bosonic fields, governed by the potential \cite{Sikivie:2006ni,Barranco:2021auj,Eby:2020ply} 
\begin{equation}\label{potential}
V(\upphi)=m_\mathfrak{a}^2 f_\mathfrak{a}^2 \left[1 - \cos\left({\upphi \over f_\mathfrak{a}} \right)\right] \,.
\end{equation}  
In this effective approach, one can consider the scenario leading to the MGD into the EKG system, implementing the energy-momentum tensor (\ref{tmunu}) together with the mean value of the 
energy-momentum tensor operator $\langle \hat T^{\mu \nu} \rangle$ associated with the quantized axion scalar field $\upphi$, with potential energy \eqref{potential}. 
 The axion decay constant $f_\mathfrak{a}$, representing the scale suppressing the effective operator, appears in the Lagrangian density regulating QCD with an 
axion field.  Denoting by $A_\mu = A^a_\mu T^a$ the $\mathfrak{su}(3)$ Lie algebra-valued gauge vector potential (for $T^a$ being the $\mathfrak{su}(3)$  generators), by $D_\mu = \partial_\mu - ig_s A_\mu^a T^a$ the covariant derivative,  by $G_{\mu\nu}^a = \partial_{[\mu}A^a_{\nu]} + g_s A_\mu^b A_\nu^c f^{abc}$ the gluon field strength in QCD, and its dual denoted by a ring, such a Lagrangian density is given by 
\begin{eqnarray}
{\cal L} &=& - {1 \over 4} G^a_{\mu\nu} G^{a\mu\nu}
+ {1 \over 2} D_\mu \mathfrak{a} D^\mu \mathfrak{a} + \sum_q \bar{q} \left(i \gamma^\mu D_\mu - m_q\right) q
+ {g_s^2 \over 32 \pi^2} \left({\mathfrak{a} \over f_\mathfrak{a}}+\theta\right) 
G^a_{\mu\nu} \mathring{G}^{a\mu\nu},
\label{QCDa}
\end{eqnarray}
where $\mathfrak{a}$ is the massless pseudoscalar
axion field and $\theta$ is a CP violating QCD angle, whereas the last term in Eq. (\ref{QCDa}) is the axion-gluon operator, regarding the effective coupling to the CP
violating topological gluon density.  Eq.~(\ref{QCDa}) uses the standard notation $g_s$ 
for  the strong coupling constant, for the quark fields, regulated by the Dirac-like Lagrangian, and their mass $m_q$. The axion decay constant is related to the magnitude $v_\mathfrak{a}$ of the VEV that breaks the U(1) symmetry in the Peccei--Quinn--Weinberg--Wilczek axion model, as $f_\mathfrak{a} = v_\mathfrak{a}/N$, for $N$ 
being an integer characterizing the U(1) color anomaly \cite{Sikivie:2006ni}.
The Lagrangian (\ref{QCDa}) describes an effective field theory, where the Standard Model can be extended by the introduction of the axion. 
The axion mass reads\footnote{See Eq. (2) of Ref. \cite{Sikivie:2006ni}. For the theoretical origin of Eq. (\ref{mass11}) in terms of the $u$ and $d$ quark masses as well as the pion mass and decay constant,  see Eq. (51) of Ref. \cite{DiLuzio:2020wdo}.}
\begin{equation}
m_\mathfrak{a} \approxeq 5.7\, \left({10^{12}~{\rm GeV} \over f_\mathfrak{a}}\right)~\mu {\rm eV}
\label{mass11}
\end{equation} is adopted, as usual. As a population of relic thermal axions was produced in the early universe, for $f_\mathfrak{a} > 10^9$ GeV, the axion lifetime exceeds by many orders of magnitude the age of the universe and the model hereon is robust for such a range of the axion decay constant. We will adopt later in Sec. \ref{main} the phenomenologically sound value $f_\mathfrak{a} \approx 10^{12}$ GeV.

The self-gravitating system arises as a solution to the EKG
equations,
\beq
\label{einstein}
\mathtt{G}_{\mu\nu}&=& \langle \hat {\mathsf{T}}_{\mu\nu} \rangle\,,\\\label{KG}
\frac{1}{\sqrt{-g}}\partial_\mu\left(\sqrt{-g}g^{\mu \nu}\partial_\nu\right)\upphi - \frac{dV(\upphi)}{d\upphi}&=&0\,,
\eeq
where the energy-momentum tensor $\mathsf{T}_{\mu\nu}$ in Eq. \eqref{einstein} encodes the $\mathtt{T}_{\mu\nu}$ tensor in Eq. (\ref{tmunu}), explicitly added with the energy-momentum tensor associated with the axion, 
\begin{equation}\label{emta}
\mathcal{T}_{\mu\nu}=g_{\mu}^{\;\rho}\partial_\rho\upphi\partial_\nu\upphi-\frac{1}{2}
\left(g^{\rho \sigma}\partial_\rho \upphi\partial_\sigma \upphi\,- V(\upphi)\right)\delta_{\mu\nu}\,,
\end{equation} 
as 
\beq
\mathsf{T}_{\mu\nu}=\mathtt{T}_{\mu\nu}+\mathcal{T}_{\mu\nu}.
\eeq
Although the first term of the energy-momentum $\mathtt{T}_{\mu\nu}$ on the right-hand side of Eq. (\ref{tmunu}) contains particles and fields on the brane, our analysis in what follows will be less intricate by considering the explicit term (\ref{emta}) summing up the axion contribution to the total energy-momentum tensor.  
With $V(\upphi)=0$, the total mass of a boson star described by the system (\ref{einstein}, \ref{KG}), ranges from 0 to a maximum of
$M_{\textsc{max}}=0.633\,m^2_p/m_{\mathfrak{a}}$, which is typically smaller than a typical neutron stellar mass.
However, if a quartic self-couple term is included,
even for a small coupling constant, the boson star mass can be comparable to a neutron star, at least in the GR case \cite{Barranco:2021auj}.

Here the MGD metric is taken into account to analyze the influence of the scalar field describing the axion in the EKG system. 
The total mass of the resulting object and the typical radius depend mainly on the 
properties of the scalar field playing the role of the axion. 
To handle the quantum nature of the axion field, the expectation value 
$\langle\hat{{\mathsf{T}}}^{\mu\nu}\rangle$ in Eq. (\ref{einstein}) must be computed -- which indeed comprises calculating  just the part 
$\langle {\hat{\mathcal{T}}}^{\mu\nu}\rangle$ in Eq. (\ref{emta}) -- 
implementing the usual quantization procedure $\upphi \mapsto \hat \upphi=\hat \upphi^++\hat \upphi^-$, where
\begin{eqnarray}\label{quantum}
\hat \upphi^{\pm}&=&\sum_{n\ell m} \mu_{n\ell m}^\pm\beta_{n\ell}(r){}^\pm Y^{\ell}_m(\theta ,\psi)e^{\mp iE_nt},\end{eqnarray} 
denoting ${}^+ Y^{\ell}_m\equiv Y^{\ell}_m$,  ${}^- Y^{\ell}_m\equiv Y^{*\ell}_m$, whereas the $\mu_{n\ell m}^{+ [-]}$ are the usual creation [annihilation] operators, with commutation relations
$\left[\mu_{n\ell m}^\pm,\mu_{n'\ell'm'}^\pm\right]=0$ and $
\left[\mu_{n\ell m}^-,\mu_{n'\ell'm'}^+\right]=-\delta_{n n'}\delta_{\ell\ell'}\delta_{m m'}$.
With the operator $\hat \upphi$, it is now possible to construct the energy-momentum tensor 
operator $\hat {\mathcal{T}}_{\mu \nu}$ just by inserting the operator $\hat \upphi$ into the formula for the energy-momentum tensor (\ref{emta}) underlying the axion field.  
The expectation value $\langle \psi\, |\hat{{\mathcal{T}}}_{\mu \nu}|\,\psi\rangle$ can be then implemented for a state $|\,\psi\,\rangle$ containing $N$ copies of the ground-state, corresponding to the $n=1$ and $\ell=0=m$ quantum numbers. 
For computing $\langle {\hat{\mathcal{T}}}^{\mu\nu}\rangle$, one performs a Taylor expansion of Eq. (\ref{potential}),  
\begin{equation}\label{taylor}
V(\upphi)=m^2\left(\frac{1}{2!}\upphi^2-\frac{1}{4!f_\mathfrak{a}^2}\upphi^4+\frac{1}{6!f_\mathfrak{a}^4}\upphi^6-\frac{1}{8!f_\mathfrak{a}^6}\upphi^8+\cdots\right)
\end{equation}
The leading self-interaction term in Eq. (\ref{taylor}) yields a $\lambda\upphi^4$-type potential, with attractive coupling $\lambda = -m_\mathfrak{a}/f_\mathfrak{a}^2$. Higher-order self-interaction
terms turn out to be relevant when high-density regimes set in \cite{Eby:2019ntd,Eby:2020ply}. 
Ref. \cite{Barranco:2021auj} showed that in the general-relativistic case, all the results for the gravitational mass, density, compactness, and radii of axion stars do not depend strongly on the number of terms considered 
in the Taylor expansion of (\ref{potential}).  
Computing the expectation value for the diagonal components of 
$\langle \hat{\mathsf{T}}_{\mu\nu}\rangle$ yields 
\begin{subequations} 
\begin{eqnarray}\label{vacuum}
\langle \hat {\mathsf{T}}^0_0\rangle&=&{\scalebox{0.98}{$\mathtt{T}$}}^0_0-\frac{E^2\beta^2}{2a}-\frac{b\beta'^2}{2}-\frac{m_\mathfrak{a}^2\beta^2}{2}
+\frac{m^2\beta^4}{12f_\mathfrak{a}^2}-\frac{m_\mathfrak{a}^2\beta^6 }{144f_\mathfrak{a}^4}+\cdots \,, \\
\langle \hat {\mathsf{T}}^1_1\rangle&=&{\scalebox{0.98}{$\mathtt{T}$}}^1_1+\frac{E^2\beta^2}{2a}+\frac{b\beta'^2}{2}-\frac{m_\mathfrak{a}^2\beta^2}{2}
+\frac{m_\mathfrak{a}^2\beta^4}{12f_\mathfrak{a}^2}-\frac{m_\mathfrak{a}^2\beta^6 }{144f_\mathfrak{a}^4}+\cdots\,, \label{vacuum1}\\
\langle \hat {\mathsf{T}}^2_2\rangle&=&{\scalebox{0.98}{$\mathtt{T}$}}^2_2+\frac{E^2\beta^2}{2a}-\frac{b\beta'^2}{2}-\frac{m_\mathfrak{a}^2\beta^2}{2}
+\frac{m^2\beta^4}{12f_\mathfrak{a}^2}-\frac{m_\mathfrak{a}^2\beta^6 }{144f_\mathfrak{a}^4}+\cdots\,,\label{vacuum2}
\end{eqnarray}
\end{subequations} 
where $\beta$ denotes $\beta_{10}$, associated with the axion ground state. In all numerical calculations that follow, the axion potential (\ref{taylor}) is expanded up to $\mathcal{O}\left(\upphi^{20}\right)$, being the error concerning the use of higher-order terms smaller than $10^{-3}\%$. 
Moreover, the terms ${\scalebox{0.98}{$\mathtt{T}$}}^0_0$, ${\scalebox{0.98}{$\mathtt{T}$}}^1_1$, and ${\scalebox{0.98}{$\mathtt{T}$}}^2_2$ in Eqs. (\ref{vacuum}) -- (\ref{vacuum2}) come from the energy-momentum tensor (\ref{tmunu}) generating the MGD solutions. When ${\scalebox{0.98}{$\mathtt{E}$}}_{\mu\nu} =0=S_{\mu\nu}$, in the absence of Kaluza--Klein modes, the GR limit is recovered. 
Considering 
 Eqs. (\ref{einstein}, \ref{KG}), with the potential (\ref{taylor}) and the static 
spherically symmetric metric 
\begin{equation}\label{abr3}
ds^{2} = \mathtt{A}(r) dt^{2} - \frac{1}{\mathtt{B}(r)}dr^{2} - r^{2}d\Upomega^2, 
\end{equation} the EKG system is obtained:
\begin{subequations}
\begin{eqnarray}\label{s1}
-\frac{\mathtt{B}'}{\mathtt{B}^2 r}+\frac{1}{r^2}\left(1-\mathtt{B}\right)&=&-\langle {\hat{\mathsf{T}}}^0_0\rangle\,, \\
\frac{\mathtt{A}\mathtt{A}'}{\mathtt{B}r}-\frac{1}{r^2}\left(1-\mathtt{B} \right)&=& \langle \hat{\mathsf{T}}^1_1\rangle \,,\label{s2}\\
\beta''+\left(\frac{2}{r}+\frac{\mathtt{A}'}{2\mathtt{A}}+\frac{\mathtt{B}'}{2\mathtt{B}}\right)\beta'+\frac{1}{\mathtt{B}}\left[\left({\mathtt{A}E^2}-\frac{m_\mathfrak{a}^2}{2}\right)\beta+
\frac{m_\mathfrak{a}^2\beta^3}{6f_\mathfrak{a}^2}-\frac{m_\mathfrak{a}^2\beta^5}{48f_\mathfrak{a}^4}\right]&=&0\,.\label{s3}
\end{eqnarray}
\end{subequations}
One can express the system
(\ref{s1}) -- (\ref{s3}) with respect to the variables $x=rm_\mathfrak{a}$, 
$\beta=4\sigma$, $\tilde{\mathtt{A}}=m_\mathfrak{a}^2\mathtt{A}/E^2$, and 
\begin{equation}\label{axionlambda}
\Lambda=\frac{m_p^2}{24\pi f_\mathfrak{a}^2}\,.
\end{equation}
 The system (\ref{s1}) -- (\ref{s3}) can be solved to constrain the axion scalar field $\beta$, with Dirichlet and Neumann conditions 
$\lim_{r\to0}\beta(r)=\beta_0$ and $\lim_{r\to0}\beta'(r)=0$. By imposing the solutions of (\ref{s1}) -- (\ref{s3}) to be regular at the origin and flat at infinity, the shooting method can be employed.
Analogously, for all figures that follow, considering the MGD parameter in Eq. (\ref{Lk}) as ${\scalebox{0.98}{$\mathfrak{l}$}}=10^{-4}\,{\rm m}$ corresponds to the brane tension $\upzeta \approx 3.118\times10^{-6} \;{\rm GeV} = 7.323 \times 10^{14}\,
{\rm kg.m^2/s^2}$, whereas ${\scalebox{0.98}{$\mathfrak{l}$}}=10^{-6}\,{\rm m}$ and ${\scalebox{0.98}{$\mathfrak{l}$}}=10^{-8}\,{\rm m}$ regard, respectively, $\upzeta \approx 3.118\times10^{-4} \;{\rm GeV}$ and $\upzeta \approx 3.118\times10^{-2} \;{\rm GeV}$. Using Eq. (\ref{uuup}), one obtains the value of the 5-dimensional Planck mass corresponding to the values of the brane tension considered here,
\beq
m_{p5} = \begin{cases}
4.8761\times 10^{-16}\,{\rm kg},&\;\;\; ({\rm for}\;\;\;{\scalebox{0.98}{$\mathfrak{l}$}}=10^{-4}\,{\rm m}),\\
1.0507\times 10^{-15}\,{\rm kg},&\;\;\; ({\rm for}\;\;\;{\scalebox{0.98}{$\mathfrak{l}$}}=10^{-6}\,{\rm m}),\\
2.2638\times 10^{-15}\,{\rm kg},&\;\;\; ({\rm for}\;\;\;{\scalebox{0.98}{$\mathfrak{l}$}}=10^{-8}\,{\rm m}).
\end{cases}
\eeq
The analysis that follows therefore takes into account how distinct ranges for the finite  brane tension can impart physical signatures on the asymptotic value of the gravitational mass, the density, the radius, and the compactness of MGD axion stars. When the brane tension increases, the results approach the GR regime of an infinitely rigid brane. 

Rewriting the metric (\ref{abr3}) in terms of $x$ and superseding them into the system (\ref{s1}) -- (\ref{s3}), one can obtain the solution for the gravitational mass function, 
as illustrated in Fig. \ref{masses1} -- \ref{masses3}, for several values of $\sigma(0)$ and $\Lambda$ (see Eq. \eqref{axionlambda}). Choosing a value of the radius $r$ which is sufficiently large, it is possible to estimate the mass $M$ of these objects as (see Eq. (3.11) of Ref. \cite{Barranco:2021auj}) as
\beq
M(x)=4\pi x \left(1- \mathtt{B}(x)\right)\frac{m_p^2}{m_\mathfrak{a}}.
\eeq
When one analyzes the asymptotic value of the mass function,
\beq\label{minfy}
{M}=\lim_{\mathtt{x}\to\infty}{M}(\mathtt{x}),\eeq 
the effective radius $R_{99}$ of a self-gravitating compact distribution defines a region that encloses 99\% of the axion star total mass, namely, $M_{99}\equiv {M}(R_{99}) = 0.99{M}$. One can emulate this concept for determining the effective radius of MGD axion stars.


\begin{figure}[H]
\centering
	\includegraphics[width=8.3cm]{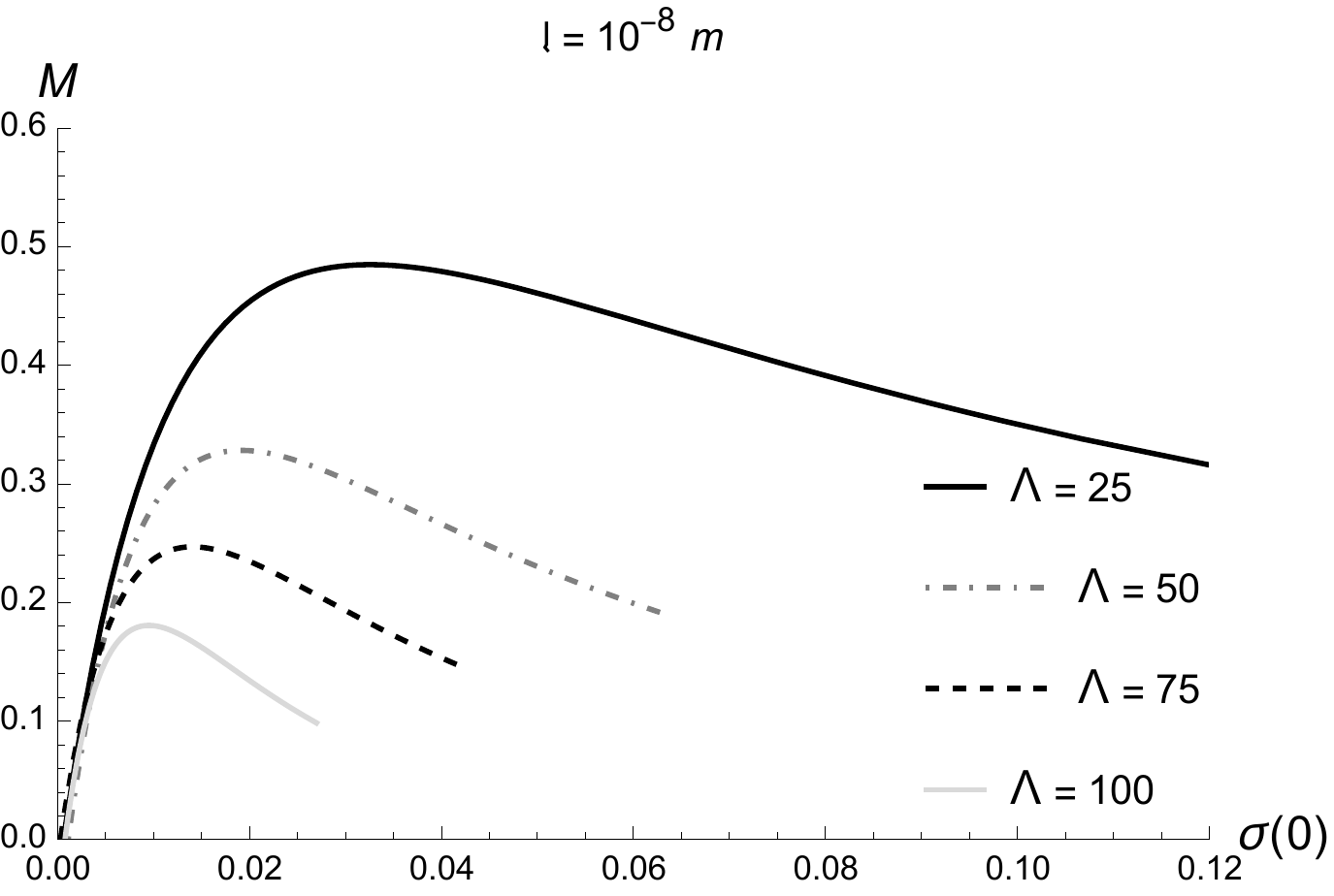}
 \caption{Asymptotic value of the gravitational mass (in units of  $m_p^2/m_\mathfrak{a}$) as a function of the central value 
of the axion scalar field $\sigma(0)$, for ${\scalebox{0.98}{$\mathfrak{l}$}}=10^{-8}\,{\rm m}$. 
The black line regards $\Lambda=25$, the dot-dashed grey line corresponds to $\Lambda=50$, the dashed black line depicts $\Lambda=75$, and the light-grey line illustrates the $\Lambda=100$ case.
 }\label{masses1}
\end{figure}


\begin{figure}[H]
\centering
	\includegraphics[width=8.3cm]{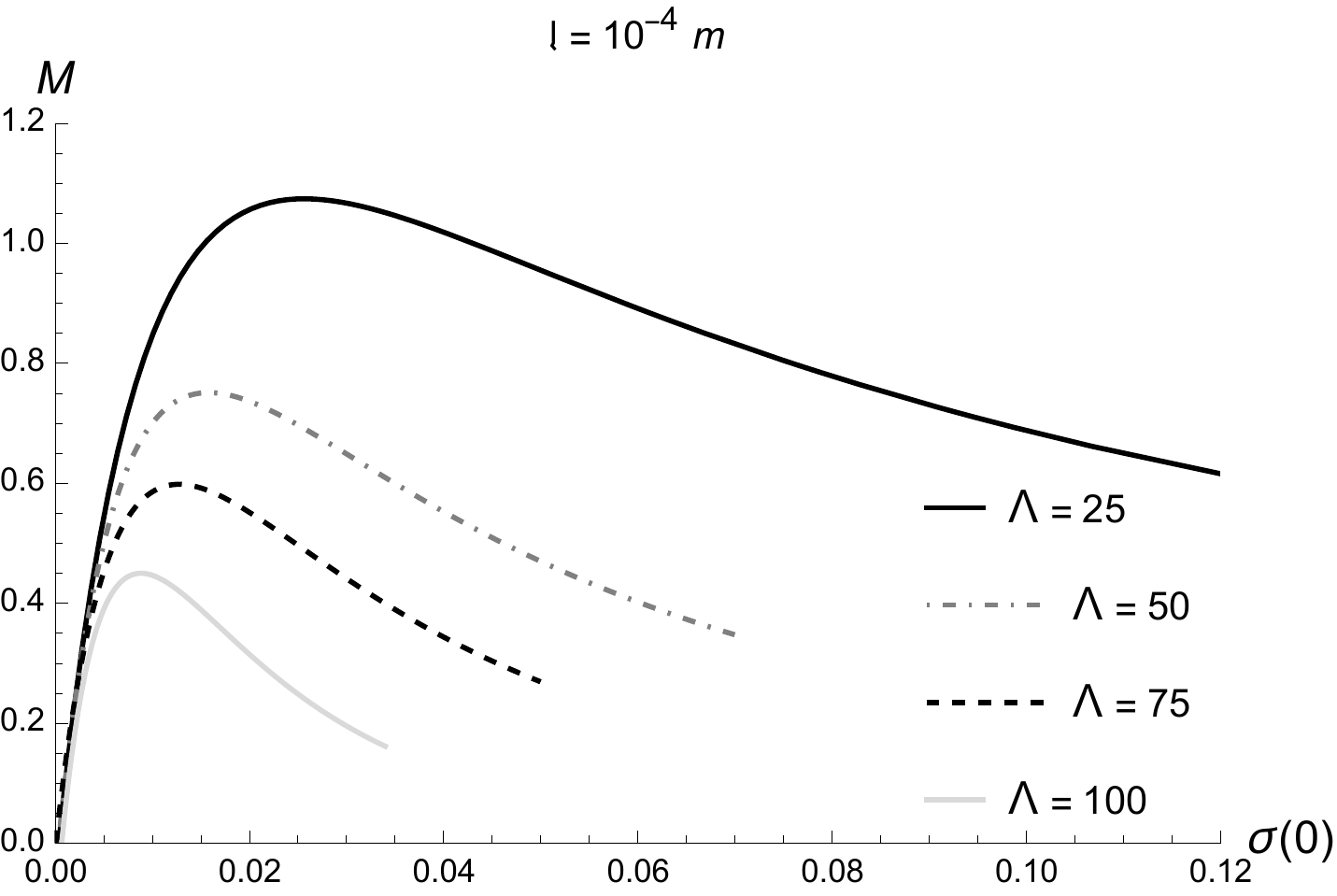}
 \caption{Asymptotic value of the gravitational mass (in units of  $m_{\textsc{Planck}}^2/m_\mathfrak{a}$) as a function of the central value 
of the axion scalar field $\sigma(0)$, for ${\scalebox{0.98}{$\mathfrak{l}$}}=10^{-4}\,{\rm m}$. 
The black line regards $\Lambda=25$, the dot-dashed grey line corresponds to $\Lambda=50$, the dashed black line depicts $\Lambda=75$, and the light-grey line illustrates the $\Lambda=100$ case.
 }\label{masses3}
\end{figure}
For each fixed value of $\sigma(0)$, the higher the value of $\Lambda$, the lower the peak $M_{\textsc{max}}$ -- denoting the maximum value of the gravitational mass function -- is. 
\begin{figure}[H]
\centering
	\includegraphics[width=8.3cm]{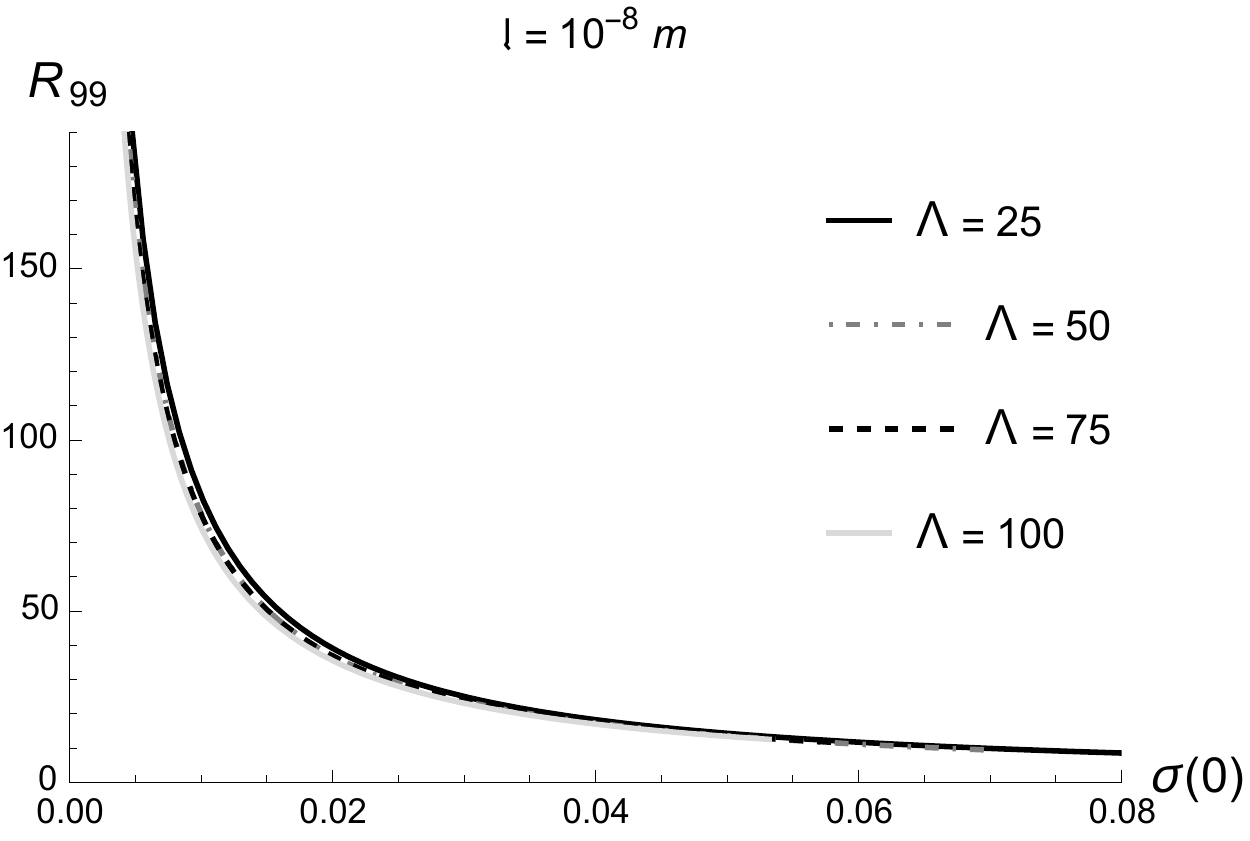}
 \caption{$R_{99}$ as a function of the central value 
of the axion scalar field $\sigma(0)$, for ${\scalebox{0.98}{$\mathfrak{l}$}}=10^{-8}\,{\rm m}$. 
The black line regards $\Lambda=25$, the dot-dashed grey line corresponds to $\Lambda=50$, the dashed black line depicts $\Lambda=75$, and the light-grey line illustrates the $\Lambda=100$ case.
 }\label{masses1}
\end{figure}


\begin{figure}[H]
\centering
	\includegraphics[width=7.8cm]{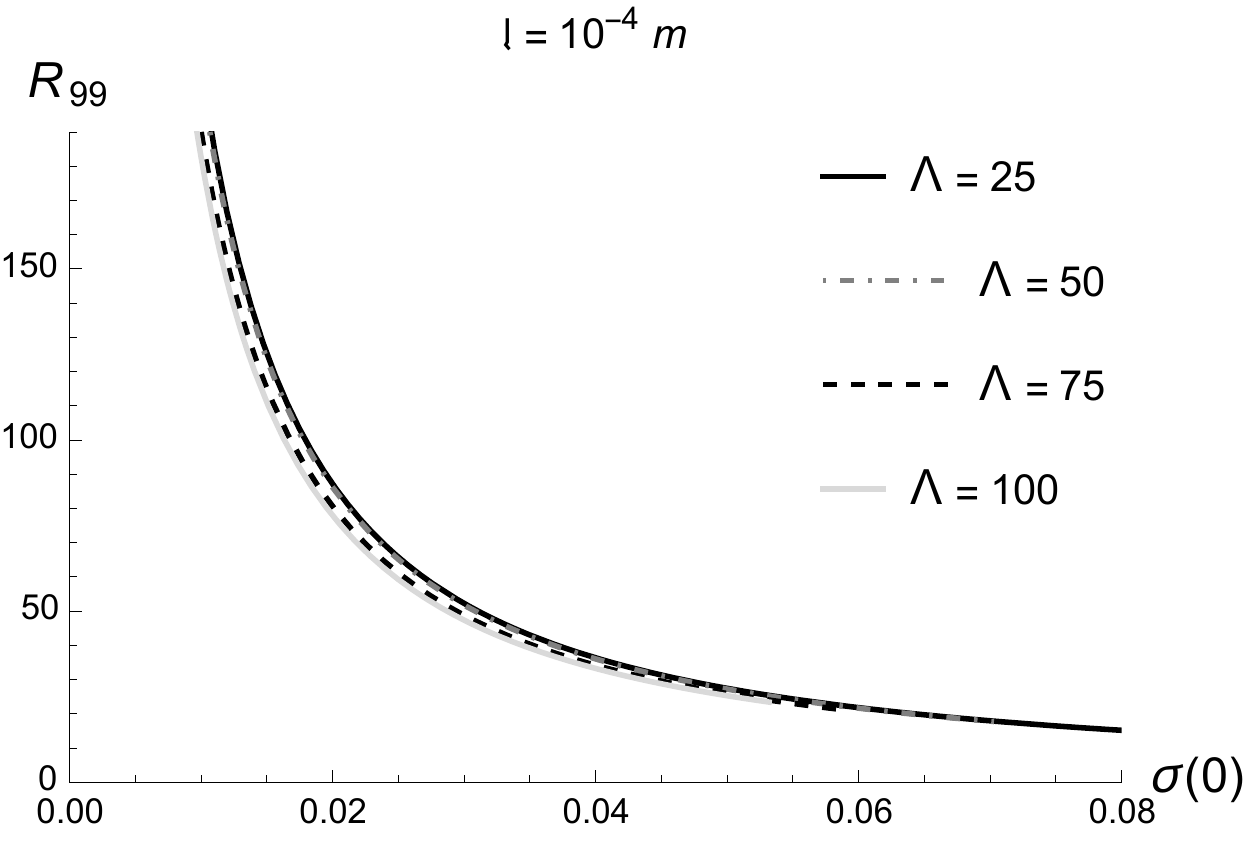}
 \caption{ $R_{99}$ as a function of the central value 
of the axion scalar field $\sigma(0)$, for ${\scalebox{0.98}{$\mathfrak{l}$}}=10^{-4}\,{\rm m}$. 
The black line regards $\Lambda=25$, the dot-dashed grey line corresponds to $\Lambda=50$, the dashed black line depicts $\Lambda=75$, and the light-grey line illustrates the $\Lambda=100$ case.
 }\label{masses3}
\end{figure}

For realistic values of the MGD parameter ${\scalebox{0.98}{$\mathfrak{l}$}}$, compatible with the physical bounds of the brane tension, the $\Lambda$-dependence of $R_{99}$ is not negligible, for lower values of $\sigma(0)$. For all values analyzed, the equilibrium configurations present a maximal mass $M_{\textsc{max}}$, at some 
value of $\sigma(0)$ that depends on the MGD parameter ${\scalebox{0.98}{$\mathfrak{l}$}}$, for each value of $\Lambda$. 
\begin{figure}[H]
\centering
\includegraphics[width=8.8cm]{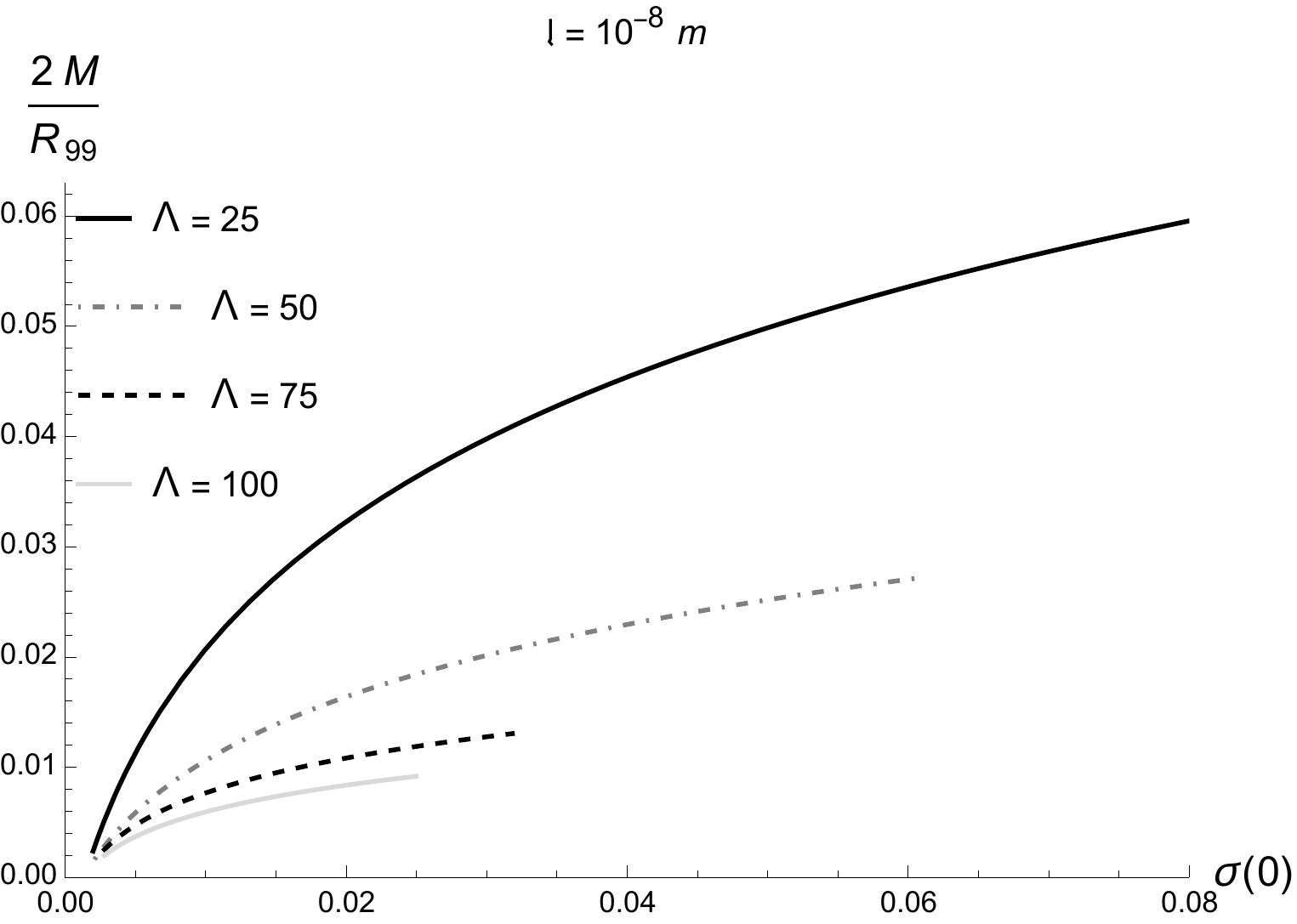}
\caption{Compactness as a function of $\sigma(0)$, for ${\scalebox{0.98}{$\mathfrak{l}$}}=10^{-8}\,{\rm m}$. 
The black line regards $\Lambda=25$, the dot-dashed grey line corresponds to $\Lambda=50$, the dashed black line depicts $\Lambda=75$, and the light-grey line illustrates the $\Lambda=100$ case.}\label{compactness1}
\end{figure}
\begin{figure}[H]
\centering
\includegraphics[width=8.8cm]{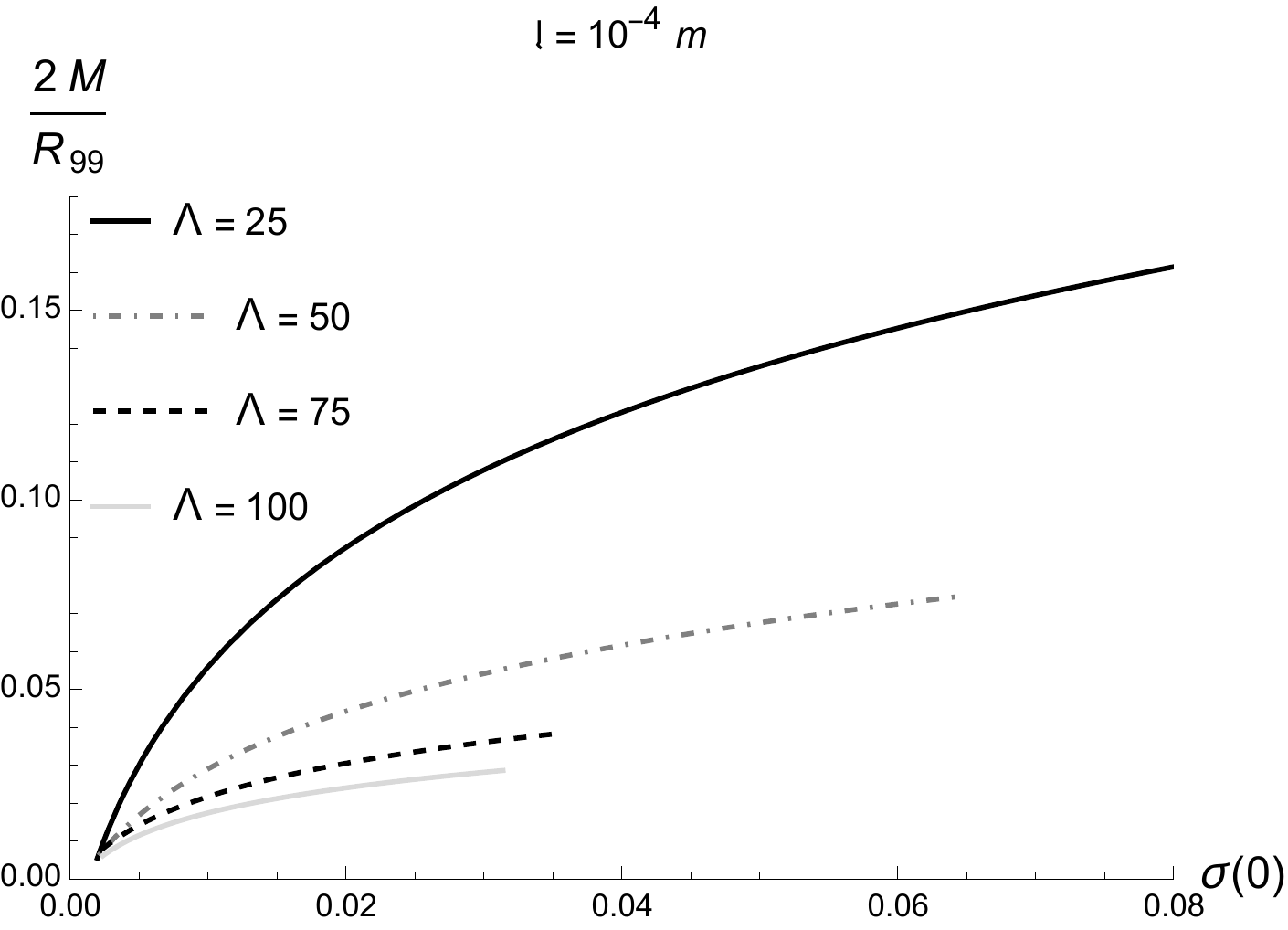}
\caption{Compactness as a function of $\sigma(0)$, for ${\scalebox{0.98}{$\mathfrak{l}$}}=10^{-4}\,{\rm m}$. 
The black line regards $\Lambda=25$, the dot-dashed grey line corresponds to $\Lambda=50$, the dashed black line depicts $\Lambda=75$, and the light-grey line illustrates the $\Lambda=100$ case.}\label{compactness3}
\end{figure}
The higher the values of $\Lambda$, the bigger the values of $M_{\textsc{max}}$ are, for each fixed value of $\sigma(0)$. The masses of equilibrium configurations, including up to the fourth power of $\upphi$ in the Taylor series, were considered in the general-relativistic limit ${\scalebox{0.98}{$\mathfrak{l}$}}\to0$ \cite{Barranco:2010ib}. 
Another interesting issue is a weak dependence of the radius $R_{99}$ on the value of $\Lambda$, irrespectively of the value of the MGD parameter, as the upper panel of Figs. \ref{masses1} and \ref{masses3} show. This feature emulates the GR limit in Ref. \cite{Barranco:2010ib}.

\section{Axion star in an MGD background}\label{main}

\begin{figure}[H]
\centering
\includegraphics[width=8.8cm]{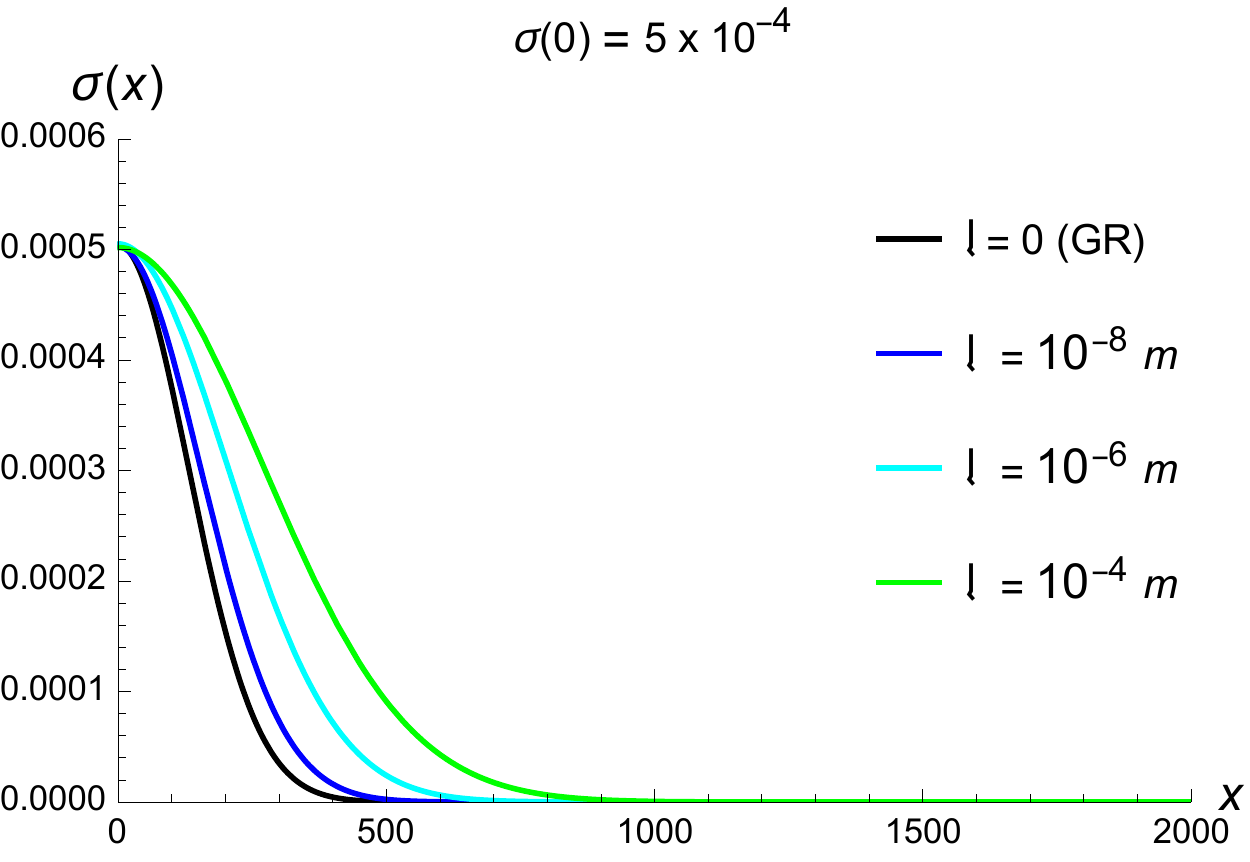}
\caption{Axion scalar field $\sigma(x)$ for a typical MGD axion stellar distribution, for 
 $\sigma(0) = 5\times 10^{-4}$. The black curve regards the general-relativistic limit ${\scalebox{0.98}{$\mathfrak{l}$}}\to0$, the blue curve illustrates the results for ${\scalebox{0.98}{$\mathfrak{l}$}}=10^{-8}$ m, and the cyan curve depicts the case where ${\scalebox{0.98}{$\mathfrak{l}$}}=10^{-6}$ m, whereas the green curve illustrates the case where ${\scalebox{0.98}{$\mathfrak{l}$}}=10^{-4}$ m.}\label{axion1}
\end{figure}

\begin{figure}[H]
\centering
\includegraphics[width=8.8cm]{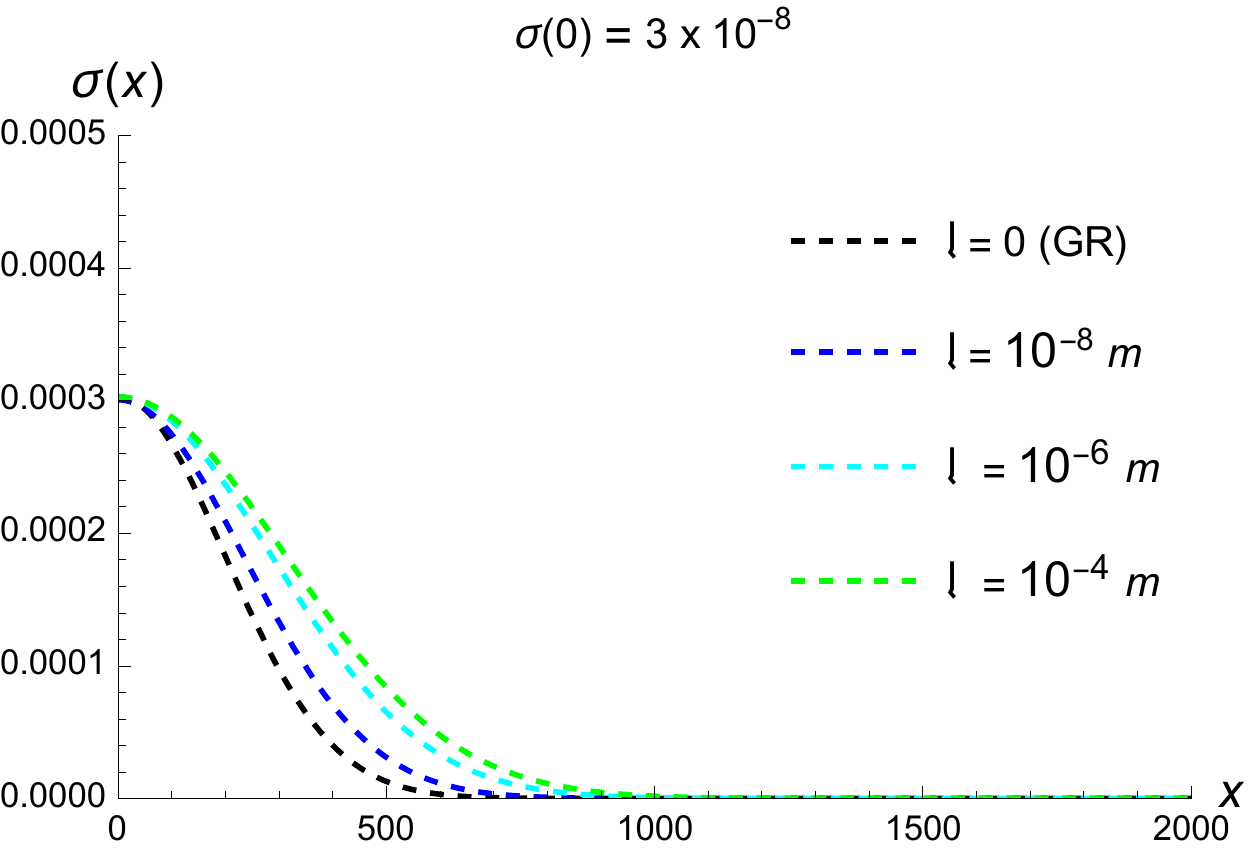}
\caption{Axion scalar field $\sigma(x)$ for a typical MGD axion stellar distribution, for 
 $\sigma(0)=3\times 10^{-4}$. The black curve regards the general-relativistic limit ${\scalebox{0.98}{$\mathfrak{l}$}}\to0$, the blue curve illustrates the results for ${\scalebox{0.98}{$\mathfrak{l}$}}=10^{-8}$ m, and the cyan curve depicts the case where ${\scalebox{0.98}{$\mathfrak{l}$}}=10^{-6}$ m, whereas the green curve illustrates the case where ${\scalebox{0.98}{$\mathfrak{l}$}}=10^{-4}$ m.}\label{axion2}
\end{figure}

\begin{figure}[H]
\centering
\includegraphics[width=8.8cm]{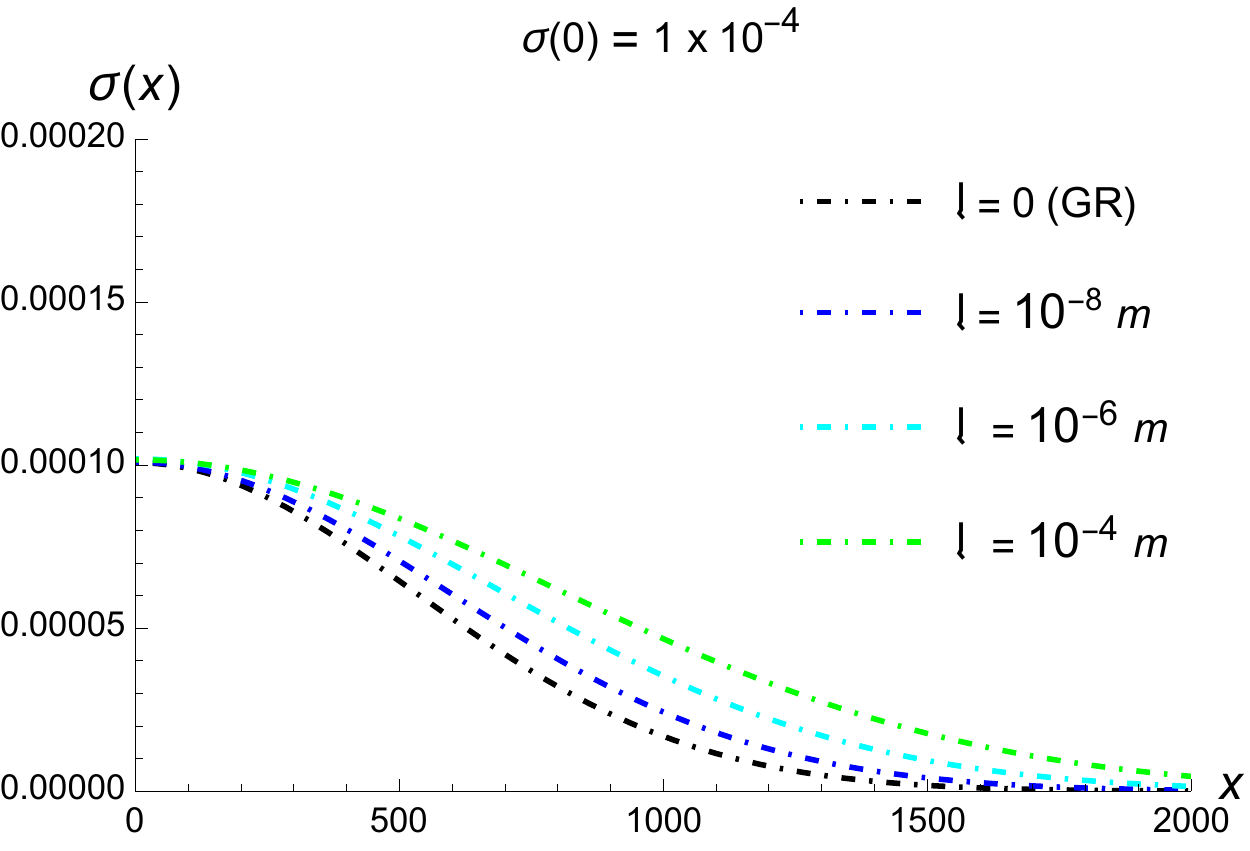}
\caption{Axion scalar field $\sigma(x)$ for a typical MGD axion stellar distribution, for 
 $\sigma(0)=1\times 10^{-4}$. The black curve regards the general-relativistic limit ${\scalebox{0.98}{$\mathfrak{l}$}}\to0$, the blue curve illustrates the results for ${\scalebox{0.98}{$\mathfrak{l}$}}=10^{-8}$ m, and the cyan curve depicts the case where ${\scalebox{0.98}{$\mathfrak{l}$}}=10^{-6}$ m, whereas the green curve illustrates the case where ${\scalebox{0.98}{$\mathfrak{l}$}}=10^{-4}$ m.}\label{axion3}
\end{figure}

After axion miniclusters are formed, the gravitational cooling effect yields some regions of the axion minicluster to become colder by ejecting axions, which leads to the formation of axion stars, with gravity balancing the quantum pressure \cite{Seidel:1993zk,Levkov:2018kau}.
All the results in Figs. \ref{axion1} -- \ref{axion3} take into account the axion mass  $m_\mathfrak{a}\approxeq 10^{-5}$ eV. In fact, regarding the Lagrangian (\ref{QCDa}), the strong CP problem can be solved as long as the vacuum energy has a minimum when the coefficient of the last term in this Lagrangian is equal to zero, making the CP-violating operator to vanish. As a consequence, the axion attains the tiny value $m_\mathfrak{a}\approxeq 10^{-5}$ eV of mass, yielding a  population of excitations in a cosmological scale,  contributing to the DM \cite{DiLuzio:2020wdo}. Regarding Figs. \ref{axion1} -- \ref{axion3}, it is worth emphasizing that the higher the value of the MGD parameter ${\scalebox{0.98}{$\mathfrak{l}$}}$, the more the axion field endures along the $x$ radial coordinate, for any value of the central value $\sigma(0)$ here analyzed. It shows that realistic values of the brane tension, encoded in the MGD parameter ${\scalebox{0.98}{$\mathfrak{l}$}}$, 
make the strength of the axion scalar field enhance, for each fixed value of $x$. 
Also, the higher the value of the MGD parameter ${\scalebox{0.98}{$\mathfrak{l}$}}$ -- correspondingly the lower the value of the brane tension -- the slower the axion scalar field $\sigma(x)$ decays along $x$. In this sense, the finite brane tension alters the kurtosis of the normal-like form of the axion field $\sigma(x)$ in Figs. \ref{axion1} -- \ref{axion3}. The general-relativistic case, ${\scalebox{0.98}{$\mathfrak{l}$}}=0$, has a mesokurtic profile, which turns into a platykurtic shape that broadens the tail of the axion scalar field $\sigma(x)$, irrespectively of the central value $\sigma(0)$ taken into account.


\begin{figure}[H]
\centering
\includegraphics[width=8.8cm]{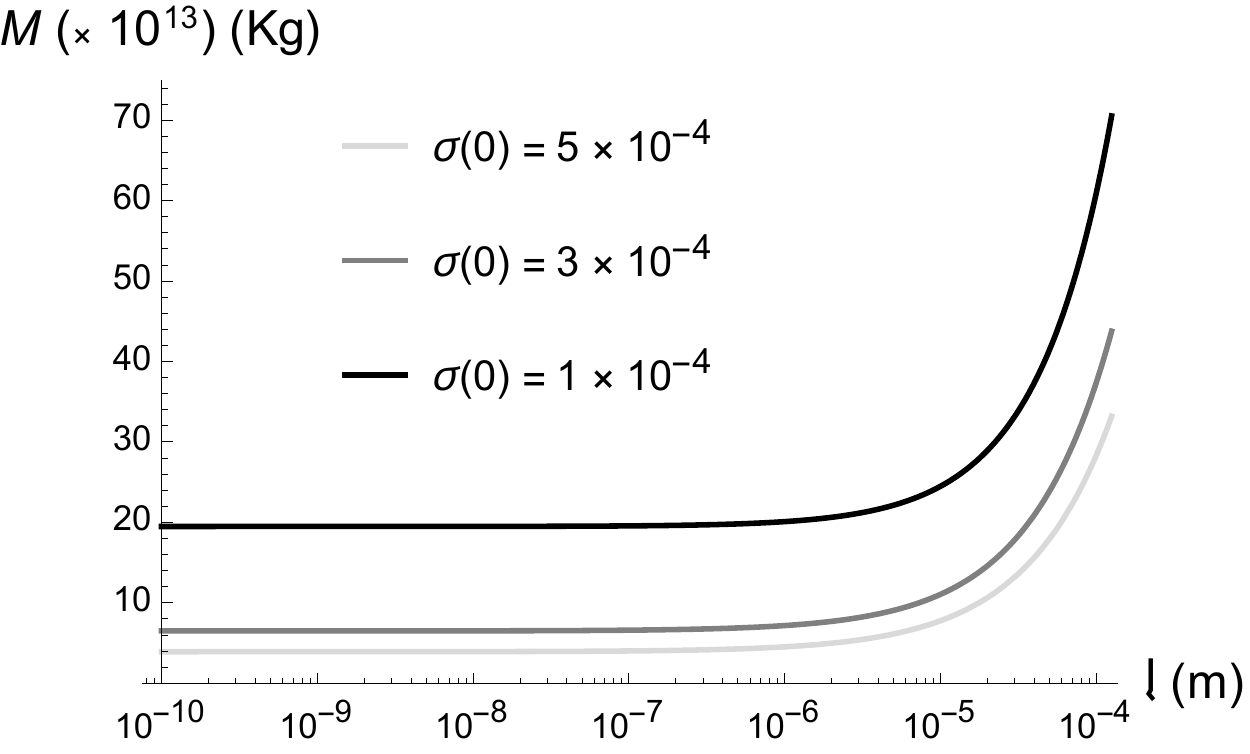}
\caption{Gravitational mass of axion star, for several values of  
 $\sigma(0)$, as a function of the MGD length ${\scalebox{0.98}{$\mathfrak{l}$}}$. The black curve 
 regards $\sigma(0) = 5\times 10^{-4}$, the grey curve illustrates the results for $\sigma(0) = 3\times 10^{-4}$ and the light-grey curve plots the case $\sigma(0) = 1\times 10^{-4}$.}\label{a1}
\end{figure}

\begin{figure}[H]
\centering
\includegraphics[width=8.8cm]{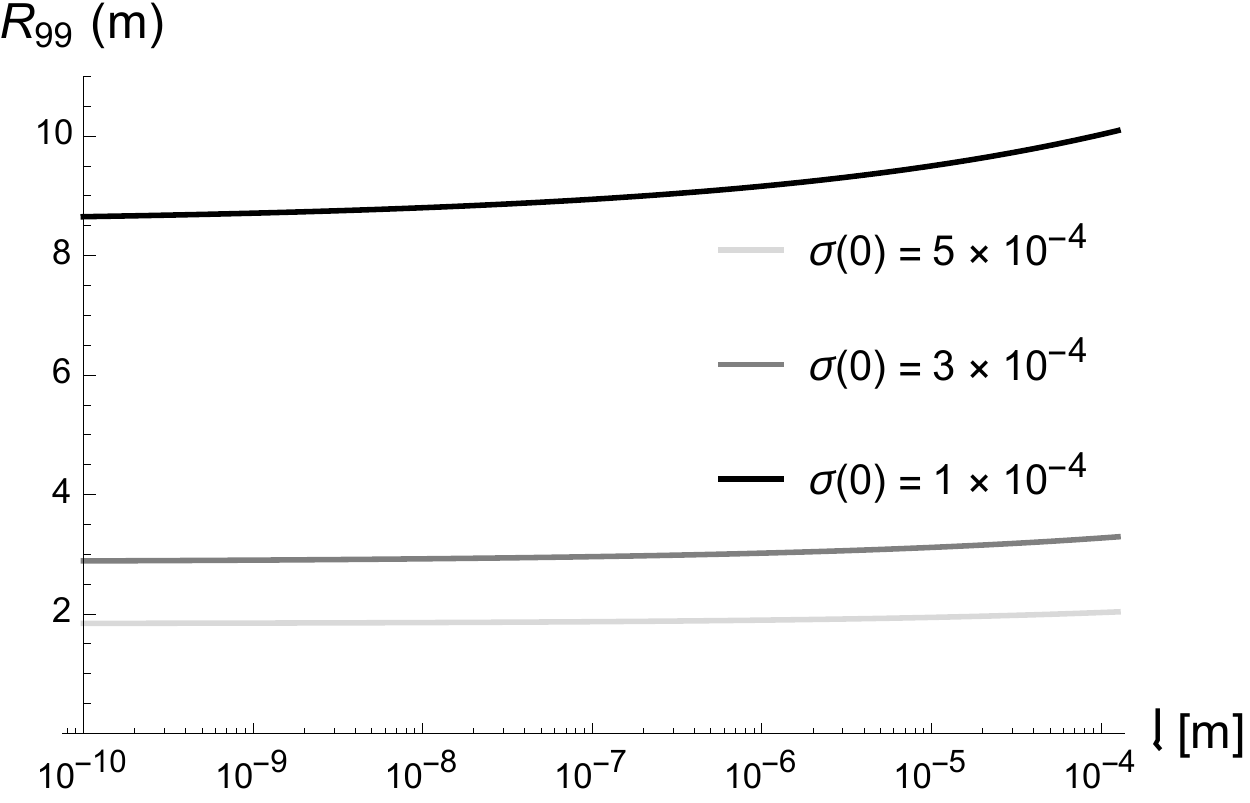}
\caption{Effective radius $R_{99}$ of axion stars, for several values of  
 $\sigma(0)$, as a function of the MGD length ${\scalebox{0.98}{$\mathfrak{l}$}}$. The black curve 
 regards $\sigma(0) = 5\times 10^{-4}$, the grey curve illustrates the results for $\sigma(0) = 3\times 10^{-4}$ and the light-grey curve plots the case $\sigma(0) = 1\times 10^{-4}$.}\label{a2}
\end{figure}

\begin{figure}[H]
\centering
\includegraphics[width=8.8cm]{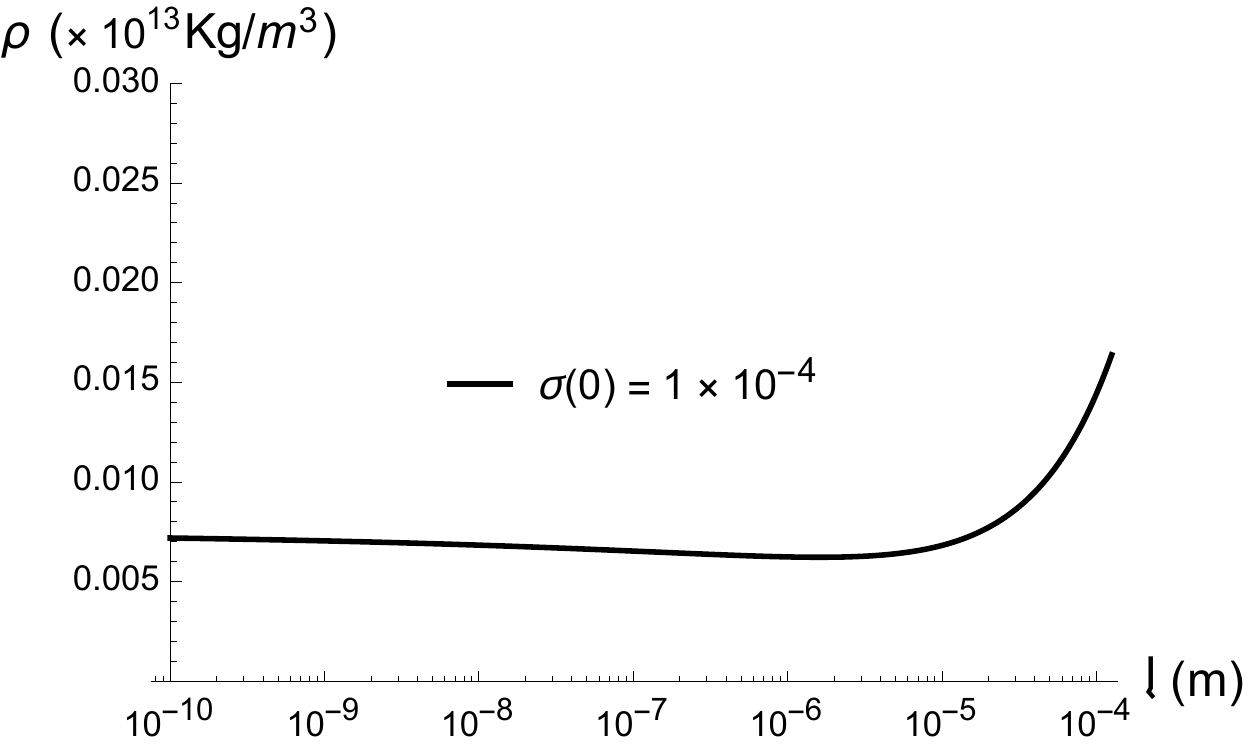}
\caption{Density of MGD axion stars, for several values of  
 $\sigma(0)$, as a function of the MGD length ${\scalebox{0.98}{$\mathfrak{l}$}}$. The black curve 
 regards $\sigma(0) = 1\times 10^{-4}$.}\label{a31}
\end{figure}

The previous results in Sec. \ref{MGD} were obtained assuming arbitrary values of the mass $m_\mathfrak{a}$
of the axions and the decay constant $f_\mathfrak{a}$.
But the mass of the axion is constrained by astrophysical and cosmological 
considerations to lie in the range $10^{-5} ~\mbox{eV} \lesssim m_\mathfrak{a} \lesssim 10^{-3}~\mbox{eV}$ and
the decay constant is related to the axion mass by Eq. (\ref{mass}) 
\cite{Sikivie:2006ni,DiLuzio:2020wdo}, yielding $10^{13}\lesssim \Lambda \lesssim10^{17}$. 
Using the variables \cite{Barranco:2010ib} 
\begin{equation}\label{variables}
\beta=\frac{f_\mathfrak{a}}{\sqrt{m}}\sigma\,, \quad r=\frac{m_p}{f_\mathfrak{a}}\sqrt{\frac{m_\mathfrak{a}}{4\pi}}x\,,\quad
{\tilde{\mathtt{A}}}=\frac{m_\mathfrak{a}^2}{E^2} \mathtt{A}\,,
\end{equation}
to solve (\ref{s1}) -- (\ref{s3}), one can realize that 
the axion star presents small compactness and low gravitational mass, for a certain range of the MGD parameter $\mathfrak{l}$. However, for higher values of the MGD parameter $\mathfrak{l}$, the MGD axion star mass increases in a steep way, as a function of $\mathfrak{l}$. 
Adopting the axion mass $m_\mathfrak{a}\approx 10^{-5}$ eV, Fig. \ref{a1} shows 
the gravitational mass of MGD axion stars, for three values of  
 $\sigma(0)$, as a function of the MGD parameter ${\scalebox{0.98}{$\mathfrak{l}$}}$. 
On the other hand, Fig. \ref{a2} illustrates the effective radius $R_{99}$ of MGD axion stars, for three values of  
 $\sigma(0)$, as a function of the MGD parameter ${\scalebox{0.98}{$\mathfrak{l}$}}$. Although the radius increases as a function of $\mathfrak{l}$, the increment  is mild
for $1\times 10^{-4}\lesssim \sigma(0) \lesssim 3\times 10^{-4}$, being a little sharper for $\sigma(0) = 5\times 10^{-4}$. 
\begin{figure}[H]
\centering
\includegraphics[width=8.8cm]{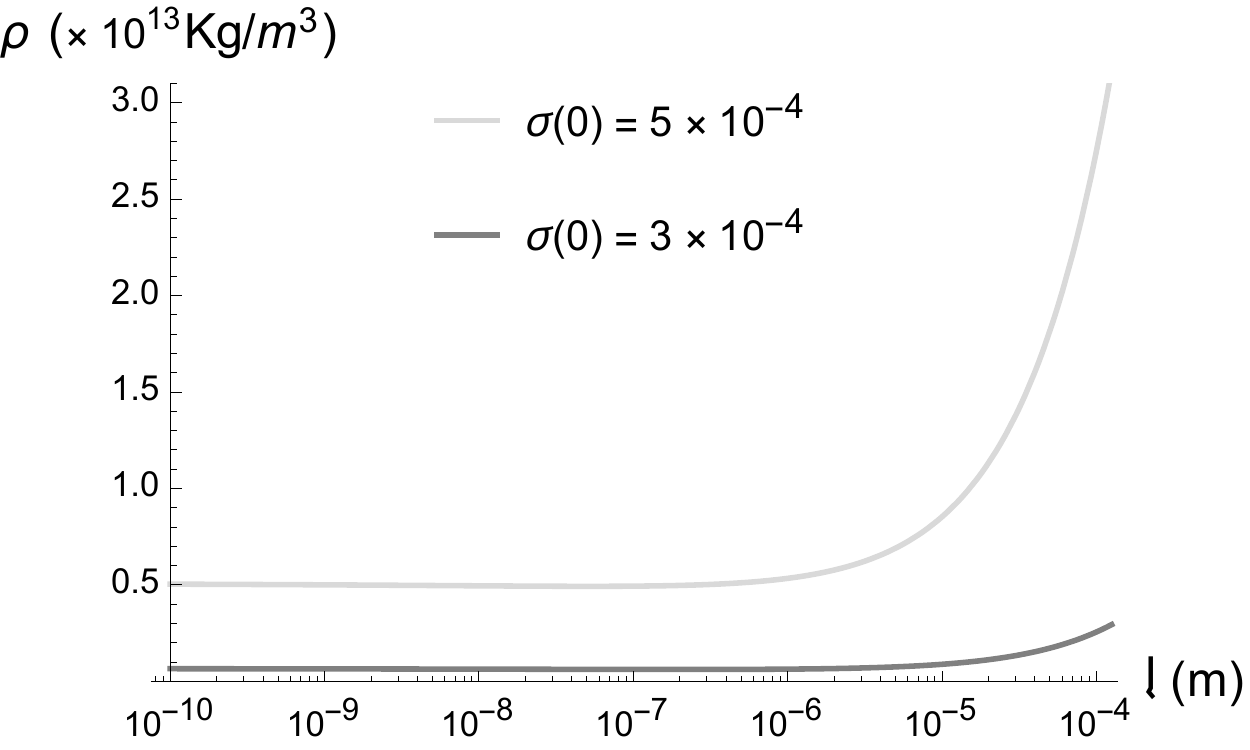}
\caption{Density of MGD axion stars, for several values of  
 $\sigma(0)$, as a function of the MGD length ${\scalebox{0.98}{$\mathfrak{l}$}}$. The grey curve illustrates the results for $\sigma(0) = 3\times 10^{-4}$ and the light-grey curve plots the case $\sigma(0) = 5\times 10^{-4}$.}\label{a3}
\end{figure}
Since the scales for the axion star density for $\sigma(0) = 5\times 10^{-4}$ differ by between 2 and 3 orders of magnitude the axion star density for $\sigma(0) = 1\times 10^{-4}$, this case is separately depicted in Fig. \ref{a31}.

The EKG system has been solved for a quantized axion scalar field governing the axion field, under the potential 
(\ref{potential}). For the MGD parameter near the general-relativistic limit, MGD axion stars have small masses and 
radii of meters, consequently having very low compactnesses. Table \ref{mass} illustrates the general-relativistic limit, matching the results in Ref. \cite{Barranco:2010ib}. 
  \begin{table}[H]
\begin{center}\medbreak
\begin{tabular}{||c||c|c|c|c||}
\hline\hline
      $\sigma(0)$ & $M$ (kg) & $R_{99}$ (m) & $\rho$ (kg/m$^3$) & \;$C={2M}/{R_{99}}$ (kg/m)\\
      \hline\hline   
      \; $5 \times 10^{-4}$\;& \;$3.903 \times 10^{13}$ \;& $1.830$ &\; $1.518 \times 10^{12}$\;&\; $4.266\times 10^{13}$ \\\hline  
      \; $3 \times 10^{-4}$ \; &\;$6.481\times 10^{13}$ \;& $2.861 $ &\;$6.613 \times 10^{11}$\;&\; $4.530\times 10^{13}$\\\hline  
     \;  $1 \times 10^{-4}$\; &\;$1.945\times 10^{14}$\; & $8.541 $ & \;$7.455 \times 10^{10}$\;&\;$ 4.554\times 10^{13}$\\
      \hline\hline  
     \end{tabular}  
\caption{Gravitational masses, $R_{99}$, density, and compactness, for axion stars in the general-relativistic limit ${\scalebox{0.98}{$\mathfrak{l}$}}\to0$ \cite{Barranco:2010ib}.}\label{mass}
\end{center}
\end{table}
In the general-relativistic limit ${\scalebox{0.98}{$\mathfrak{l}$}}\to0$, the gravitational mass of axion stars, their radius $R_{99}$, and corresponding density, for several values of $\sigma(0)$, are shown in Table \ref{mass}. Using these values, their compactness, $C=2M/R_{99}$ can be read off, lying in the range 
$10^{13} - 10^{14}$ kg/m. Since the compactness of the Sun is given by $5.71798\times 10^{21}$ kg/m, the compactness of axion stars 
equals between 7 and 8 orders of magnitude smaller than the Solar compactness. 
The MGD axion star has typical asteroid-size masses, $M\approx 10^{-17} - 10^{-16} M_{\odot}$, for ${\scalebox{0.98}{$\mathfrak{l}$}}\lesssim 10^{-5}$ m. If DM is mainly constituted by axions, the axion field might have evolved in the early universe, originating axion miniclusters. 
These structures can relax by gravitational cooling, evolving to boson stars made of axions \cite{Seidel:1993zk}. Gravitational cooling
ends in a unique final state independent of the initial conditions. 
One can realize that the typical densities for axion stars, in the general-relativistic limit, illustrated in Table \ref{mass}, lies between 5 and 7 orders of magnitude smaller than neutron star density, with average density $3.7\times 10^{17}$ to $5.9\times 10^{17}$ kg/m${}^3$, respectively corresponding to $2.6\times 10^{14}\rho_\odot$ to $4.1\times 10^{14} \rho_\odot$. 

It is already known that strong gravitational lensing effects set up the bound range $|{\scalebox{0.98}{$\mathfrak{l}$}}|\lesssim6.370\times 10^{-2}\ {\rm m}$ \cite{Cavalcanti:2016mbe}. 
Taking the upper bound of this limit yields the values of gravitational mass, $R_{99}$, density, and compactness, for several values of $\sigma(0)$, displayed in Table \ref{masslimit}.
  \begin{table}[H]\label{masslimit}
\begin{center}\medbreak
\begin{tabular}{||c||c|c|c|c||}
\hline\hline
      $\sigma(0)$ & Mass (kg) & $R_{99}$ (m) & $\rho$ (kg/m$^3$)& \;$C={2M}/{R_{99}}$ (kg/m) \\
      \hline \hline   
      \; $5 \times 10^{-4}$\;& \;$4.431 \times 10^{17}$\; & $2.780$ & $1.667 \times 10^{16}$\,&\;$3.187\times 10^{17}$\\ \hline
      \; $3 \times 10^{-4}$ \; &\;$6.288\times 10^{17}$\; & $4.410 $ &$1.805 \times 10^{15}$\,&\;$2.851\times 10^{17}$ \\ \hline
      \;$1 \times 10^{-4}$\; &\;$1.504\times 10^{18}$\; & $13.865 $ & $1.514 \times 10^{14}$\,&$2.169\times 10^{17} $\\ \hline
      \hline
     \end{tabular}  
     \caption{Gravitational masses, $R_{99}$, density, and compactness, for axion stars in the MGD background, in the extremal upper limit ${\scalebox{0.98}{$\mathfrak{l}$}}= 6.370\times 10^{-2}\ {\rm m}$ \cite{Cavalcanti:2016mbe}.}\label{masslimit}
\end{center}
\end{table}
For the MGD parameter far from the general-relativistic limit, MGD axion stars have bigger masses, being 4 orders of magnitude more massive axion stars in the general-relativistic limit. Their radii are still bigger, however still around the same order of magnitude, having still the order of meters. Consequently, MGD axion stars have still low compactnesses when compared to the Sun, although they are 4 orders of magnitude larger than axion stars in the general-relativistic limit. 
For $\sigma(0)=5 \times 10^{-4}$, MGD axion stars have a density of 1 order of magnitude smaller than neutron stars. This value for the MGD axion star density and gravitational mass makes it more difficult to be disrupted by tidal forces, when colliding near neutron stars, increasing the Roche radius. Considering a neutron star of mass $M_{\textsc{n}}$, for the MGD axion star with mass $M$ and radius $R_{99}$ to undergo tidal disruption effects, the tidal forces that act on it must have the same order of magnitude of the forces that keep the star cohesive. Estimating these forces, the maximum distance $r_{\textsc{max}}$ that allows the MGD axion star to undergo a tidal disruption event is given by \cite{Teukolsky:1974yv}
\beq
r_{\textsc{max}} = \sqrt[3]{\frac{M_{\textsc{n}}}{M}}R_{99}.
\eeq
Therefore one can plot the maximum distance $r_{\textsc{max}}$ as a function of the parameter $\mathfrak{l}$, for the three values of $\sigma(0)$ up to here analyzed. 
\begin{figure}[H]
\centering
\includegraphics[width=8.8cm]{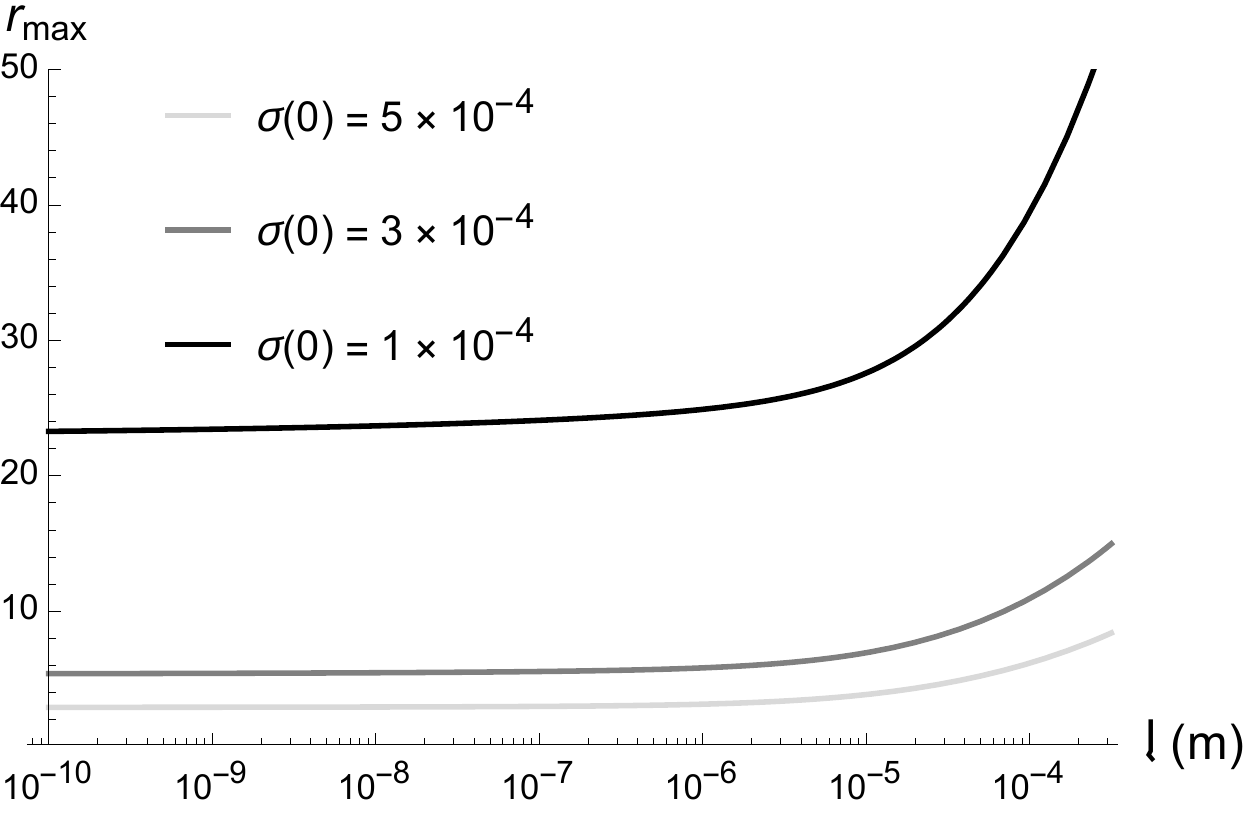}
\caption{Maximum distance $r_{\textsc{max}}$ (in units of $\sqrt[3]{M_{\textsc{n}}}$) as a function of the parameter $\mathfrak{l}$. 
The black curve indicates $\sigma(0) = 1\times 10^{-4}$, the grey curve regards $\sigma(0) = 3\times 10^{-4}$ and the light-grey curve illustrates the results for $\sigma(0) = 5\times 10^{-4}$.}\label{rm11}
\end{figure}
Fig. \ref{rm11} shows that for $\sigma(0)=1\times 10^{-4}$, the maximum distance $r_{\textsc{max}}$ increases softly as a function of $\mathfrak{l}$ up to $\mathfrak{l}\lesssim 10^{-6}$ m, which becomes a sharper dependence $r_{\textsc{max}}(\mathfrak{l})\propto (\log_{10}\mathfrak{l})^{\,3.82}$, for $\mathfrak{l}\gtrsim 5\times 10^{-4}$ m. Now, for $\sigma(0)=3\times 10^{-4}$,  $r_{\textsc{max}}$ increases nearly constant  as a function of the MGD parameter $\mathfrak{l}$, up to $\mathfrak{l}\lesssim 10^{-6}$ m, turning to a steeper dependence $r_{\textsc{max}}(\mathfrak{l})\propto (\log_{10}\mathfrak{l})^{\,1.98}$, for $\mathfrak{l}\gtrsim 4\times 10^{-5}$ m. The last case comprises $\sigma(0)=5\times 10^{-4}$, for which 
the radius $r_{\textsc{max}}$ increases nearly constant  as a function of $\mathfrak{l}$ up to values approaching $\mathfrak{l}\lesssim 10^{-5}$ m, having a sharper  dependence $r_{\textsc{max}}(\mathfrak{l})\propto (\log_{10}\mathfrak{l})^{\,1.52}$, for $\mathfrak{l}\gtrsim 2\times 10^{-5}$ m. The lower the value brane tension -- corresponding to higher values of the MGD parameter $\mathfrak{l}$ -- the larger the maximum distance $r_{\textsc{max}}$ is, permitting the MGD axion star to undergo a tidal disruption event. Therefore MGD axion stars are less sensitive to tidal disruption effects, as $\mathfrak{l}$ increases. It also corroborates the fact that their density increases as $\mathfrak{l}$ increases. Denser compact objects are more cohesive and less inclined to tidal disruption than their GR counterparts. 
MGD axion stars are even more robust to tidal disruption events for lower values of the brane tension, specifically for $\mathfrak{l}\gtrsim 10^{-6}$.

It is worth pointing out that exclusively for the case $\sigma(0)=5 \times 10^{-4}$, when the MGD parameter lies in the tiny range ${\scalebox{0.98}{$\mathfrak{l}$}}\gtrsim 9.84 \times 10^{-3}$ m, the axion field typical densities can induce stimulated decays of the axion to photons \cite{Carvalho:2022mgk}. Axion miniclusters have a standard density equal to $\approx 10^{10}$ kg/m$^3$, at which the
annihilation $\mathfrak{a}\mathfrak{a} \rightarrow \gamma\gamma$, including other eventual dissipative processes, are not importantly effective. Hence axion miniclusters undergo collapsing due to gravitational cooling, after separating from the motion of galaxies due solely to the expansion of the Universe, which characterizes the Hubble flow. Regarding axions with mass
$m\approx 10^{-5}$~eV, the maximum axion star mass equals 
$\approx 10^{25}$~kg, representing a bigger amount than the minicluster mass \cite{Seidel:1993zk}.
Hence one might expect the collapse to yield an axion star, with density $\rho \approx 10^{15}$~kg/m$^3$. However, at these densities, stimulated
decay of axions begins to be relevant, as the axion decay rate is too small, of order $\approx 10^{-49}$ sec$^{-1}$, for
$m_a\approx 10^{-5}$~eV.
The amplification arising from the stimulated decay of axions into photons yields a factor 
$\exp(D)$, with 
\begin{equation}
 D \approx {\Gamma_\pi\, m_p^2\, f_\pi V_{\textsc{esc}} \over m_\pi^4 f_\mathfrak{a}\,R}
\eqnum{2}
\end{equation}
where $\Gamma_\pi \approx 8$~eV, $f_\pi \approx 134 $~MeV,
$f_a\approx 10^{12}$~GeV, 
$m_\pi =134.977$~MeV, $V_{\textsc{esc}}=\sqrt{{2GM\over R}}$ is the escape velocity.
For MGD axion stars with minicluster mass, $D\approx 10^{27}$. It implies that the final stage of the collapse process induced by gravitational cooling is a flash, comprising a bright beam of photons \cite{Seidel:1993zk,Tkachev:1987cd}. This possibility can be traced by ground-based telescopes. 
This case does not occur in the GR-limit, as one can check the highest possible densities for MGD axion stars in Table \ref{mass}. Now, for the cases $\sigma(0)=3 \times 10^{-4}$ and $\sigma(0)=5 \times 10^{-4}$, when the MGD parameter lies in the tiny range ${\scalebox{0.98}{$\mathfrak{l}$}}\gtrsim 9.84 \times 10^{-3}$ m, the axion densities induces stimulated decays of the axion to photons. More precisely, for any value of $\sigma(0)\lesssim 2.932 \times 10^{-4}$, whatever the value of the MGD parameter is, there will be no stimulated decays of the axion to photons, and axions are a DM candidate. The axion field can form compact self-gravitating objects if $\sigma(0)\lesssim 2.932 \times 10^{-4}$, for any value of the MGD parameter. For values $\sigma(0)\gtrsim 2.932 \times 10^{-4}$, the MGD parameter must be in the tiny range ${\scalebox{0.98}{$\mathfrak{l}$}}\gtrsim 9.84 \times 10^{-3}$ m, for stimulated decays of axions to be observed.

Typical densities for axion stars are also shown in Table \ref{mass}, for the GR-limit, and in Table \ref{masslimit}, for the extremal upper limit ${\scalebox{0.98}{$\mathfrak{l}$}}= 6.370\times 10^{-2}\ {\rm m}$ \cite{Cavalcanti:2016mbe}. 
Due to the smallness of the axion star masses, the MGD axion stars can play the role of the mini-massive compact halo objects, composed by condensation of axion field, representing the final state of axion miniclusters originated in the QCD epoch of the universe evolution \cite{Witte:2022cjj}. MGD axion stars comprise a large number of stable asteroid-sized 
scalar condensations, whose final stage encompasses clustering into typical structures that are similar to cold DM halos. Assuming that the axion is the main component of DM, 
the galactic halo can be modeled by an ensemble of MGD axion stars.  
For $\sigma(0)=5 \times 10^{-4}$, the MGD axion star mass lies in the range 
\beq
1.962\times 10^{-17} M_\odot\lesssim M\lesssim 2.228\times 10^{-13} M_\odot.
\eeq
The lower limit corresponds to the GR limit ${\scalebox{0.98}{$\mathfrak{l}$}}=0$, as in Table \ref{mass}, whereas the higher limit regards the observational upper limit ${\scalebox{0.98}{$\mathfrak{l}$}}= 6.370\times 10^{-2}\ {\rm m}$ \cite{Cavalcanti:2016mbe}. 
Also, considering the same extremal limits for ${\scalebox{0.98}{$\mathfrak{l}$}}$, for $\sigma(0)=3 \times 10^{-4}$, the MGD axion star mass lies in the range 
\beq
2.784\times 10^{-17} M_\odot\lesssim M\lesssim 3.162\times 10^{-13} M_\odot,
\eeq
whereas for $\sigma(0)=1 \times 10^{-4}$, the MGD axion star mass is in the range 
\beq
6.658\times 10^{-17} M_\odot\lesssim M\lesssim 7.652\times 10^{-13} M_\odot,
\eeq

\section{Conclusions and perspectives}\label{cppp}

We showed that MGD axion stars have the asymptotic value of gravitational masses, the radii, the densities, and the compactnesses variable, expressed as a function of the brane tension. More specifically, MGD axion stars present 
typical masses and densities that can reach 4 orders of magnitude larger than GR axion stars, for a given range of brane tension.  Several other physical features of MGD axion stars were addressed, with important corrections to the general-relativistic limit.  When realistic values of the brane tension are taken into account,  
the strength of the axion scalar field enhances along the radial coordinate. 
MGD axion stars have typical masses and densities that make them less sensitive to tidal disruption, in collisions with neutron stars, for a certain range of the brane tension. The maximum distance beyond which MGD axion stars undergo tidal disruptive events was computed, for several values of the central value of the axion field, and was shown to be an increasing function of the MGD parameter, which is inversely proportional to the fluid brane tension. With it, we show that MGD axion stars are less  sensitive to tidal disruption effects, as the brane tension decreases. 

The collapse of MGD axion stars can further play the role of an important ingredient in the formation of the recently observed black holes of a nearly solar mass, which 
cannot be explained by usual theories of black hole formation \cite{LIGOScientific:2020zkf}. According to the values of the axion decay constant $f_\mathfrak{a}$ here used, the final stage of the collapse of axion stars can 
correspond to black holes. For the extremal upper limit ${\scalebox{0.98}{$\mathfrak{l}$}}= 6.370\times 10^{-2}\ {\rm m}$ \cite{Cavalcanti:2016mbe}, and for the case $\sigma(0)=5 \times 10^{-4}$, MGD axion stars have a density 1 order of magnitude smaller than neutron stars, being possible to constitute a binary system. GWs originated from the merging process coalescing binaries of MGD of compact stars, which might have a  ringdown phase after merging. In the brane-world scenario of a compact extra dimension, GWs are expected to be detected in a range of frequencies that are considerably higher than the $\sim 10^4$ Hz \cite{Chrysostomou:2022evl,Bailes:2021tot}. 
Therefore the quasinormal ringing signatures in GWs emitted from MGD axion star binaries will be essentially unique and potentially detectable and observed in ground-based telescopes \cite{Yu:2019jlb}. 
Some other aspects, including the instability and turbulence underlying solutions of Einstein's field equations coupled to the axion field, can be investigated using the apparatus developed in Ref. \cite{Barreto:2022ohl}.
In the collision process with neutron stars, photons can be emitted in the collision process with axions. If the photon plasma surrounding the neutron star has the same order of magnitude as the MGD axion mass, the axion conversion into a photon is a  coherent source, having typical radio-wave frequencies to be detected by ground-based telescopes. We also studied the tidal forces in the collision process of MGD axion stars to neutron stars. The maximum distance 
for which the MGD axion star undergoes tidal disruption event and the percentage of axions that can be converted into photons, across the collision event to neutron stars, was shown to increase as a function of the MGD parameter, corresponding to lower values of the brane tension. When one takes into account phenomenologically feasible values of the axion mass and the axion decay constant, for some range of the brane tension stimulated decay of axions into photons does occur, implying that the final stage of the collapse process induced by gravitational cooling is a flash of photons. This phenomenon has no analogy for axion stars in the general-relativistic limit, due to their lower typical densities.

We are currently in an unparalleled position wherein one can observe gravitational radiation. The LIGO--Virgo--KAGRA collaboration has validated ninety GW events with a sound probability of astrophysical source \cite{LIGOScientific:2020zkf}. It provides a unique opportunity to test extensions of GR in the strong-field regime and extensions, as the MGD solutions, in this fruitful era of GW astronomy. The range of mass for MGD axion stars, $10^{-17} M_\odot\lesssim M\lesssim 10^{-13} M_\odot$, characterizes a diluted axion star, with self-gravity and quantum pressure can be neglected compared to the gravitational force from a gravitationally bound neutron star \cite{Bai:2021nrs}.    One can use the results here obtained to test collisions 
between MGD axion stars and neutron stars. Since GWs interact weakly across their propagation, it can eventually provide relevant signatures of the inflationary epoch \cite{Chrysostomou:2022evl}.
Since the gravitational mass and the density of MGD axion stars were here shown to have 4 orders of magnitude larger than the GR, being their disruption from tidal forces under collision with neutron stars less feasible, it can provide unique observational signatures.

\subsection*{Acknowledgements}
R.C. is partially supported by the INFN grant FLAG and his work has also been carried out in the framework of activities of the National Group of Mathematical Physics (GNFM, INdAM).  
R.dR. thanks to the São Paulo Research Foundation (FAPESP) (grants No. 2022/01734-7 and No. 2021/01089-1); the  National Council for Scientific and Technological Development -- CNPq (grant No. 303390/2019-0); and the Visiting Researcher Position at DIFA (Prot. no. 002828 del 21/12/2022 - Contratto 100/2022), for partial financial support. R.dR thanks R.C. and DIFA - UniBo, for the hospitality. 
\bibliography{bib_DSS}\end{document}